\documentclass[aps,preprint,groupedaddress,showpacs]{revtex4}

\begin{document}
\title{Energetic Consistency and Momentum Conservation in the
Gyrokinetic Description of Tokamak Plasmas}
\author{B. Scott}
\email[email: ]{bds@ipp.mpg.de}
\homepage[\\ URL: ]{http://www.rzg.mpg.de/~bds/}
\author{J. Smirnov}
\affiliation{Max-Planck-Institut f\"ur Plasmaphysik, 
		Euratom Association,
                D-85748 Garching, Germany}

\date{\today}

\begin{abstract}
Gyrokinetic field theory is addressed in the context of a general
Hamiltonian.  The background magnetic geometry is static and
axisymmetric, and all dependence of the Lagrangian upon dynamical
variables is in the Hamiltonian or in free field terms.   Equations for
the fields are given by functional derivatives.  The symmetry through
the Hamiltonian with time and toroidal angle invariance of the geometry
lead to energy and toroidal momentum conservation.  In various levels of
ordering against fluctuation amplitude, energetic consistency is exact.
The role of this in underpinning of conservation laws is emphasised.
Local transport equations for the vorticity, toroidal momentum, and
energy are derived.  In particular, the momentum equation is shown for
any form of Hamiltonian to be well behaved and to relax to its
magnetohydrodynamic (MHD) form when long wavelength approximations are
taken in the Hamiltonian.
Several currently used forms, those which form the basis of most global
simulations, are shown to be well defined within the
gyrokinetic field theory and energetic consistency.
\end{abstract}


\maketitle

\def\emskip{\hskip 1 em}
\def\hfb{\hfil\break}
\def\etc{{\it etc.}}
\def\visavis{{\it vis-a-vis}\ }
\def\ie{{\it i.e.}}
\def\eg{{\it e.g.}}
\def\etal{{\it et al}}
\def\ua{u.a.\ }
\def\dh{d.h.\ }
\def\zb{z.B.\ }
\def\bzw{bzw.\ }
\def\usw{usw.\ }

\def\idelta{$i$-delta}


\def\half{ {1\over 2} }
\def\third{ {1\over 3} }
\def\fourth{ {1\over 4} }
\def\tth{ {2\over 3} }
\def\twothirds{ {2\over 3} }
\def\threehalves{ {3\over 2} }
\def\fivehalves{ {5\over 2} }
\def\fivethirds{ {5\over 3} }
\def\sevenhalves{ {7\over 2} }
\def\threeh{\threehalves}
\def\eps{\epsilon}
 
\def\grapprox{\mathop{\lower.5ex \hbox{$\buildrel{\fivesy >}\over{\fivesy\sim}$}} \nolimits}
\def\lsapprox{\mathop{\lower.5ex \hbox{$\buildrel{\fivesy <}\over{\fivesy\sim}$}} \nolimits}
\def\grls{\mathop{\lower.5ex \hbox{$\buildrel{\fivesy >}\over{\fivesy <}$}} \nolimits}

\def\vec#1{{\bf #1}}
\def\tsr#1{{\secfnt #1}}
\def\avg#1{\left\langle #1 \right\rangle}
\def\abs#1{\left\vert #1 \right\vert}
\def\prf#1{\overline{#1}}

\def\max{{}_{{\rm max}}}
\def\min{{}_{{\rm min}}}

\def\minus{\mathop{\hbox{--}}\nolimits}

\def\re{\mathop{\rm Re}\nolimits}
\def\im{\mathop{\rm Im}\nolimits}
\def\csch{\mathop{\rm csch}\nolimits}
\def\sech{\mathop{\rm sech}\nolimits}
\def\diag{\mathop{\rm diag}\nolimits}
\def\Max{\mathop{\rm Max}\nolimits}
\def\Min{\mathop{\rm Min}\nolimits}
\def\nint{\mathop{\rm NINT}\nolimits}
\def\mod{\mathop{\rm mod}\nolimits}
\def\det{\mathop{\rm det}\nolimits}
\def\Tr{\mathop{\rm Tr}\nolimits}
\def\sign{\mathop{\rm sign}\nolimits}

\def\LBR{\left\lbrace}
\def\RBR{\right\rbrace}
\def\LB{\left\lbrack}
\def\RB{\right\rbrack}
\def\LP{\left (}
\def\RP{\right )}
\def\qqquad{\qquad\qquad}
\def\qqqquad{\qquad\qquad\qquad}
\def\Det#1{\left\vert\matrix{#1}\right\vert}

\def\pt{\partial}

\def\pzz#1{{\partial #1\over\partial z}}
\def\pxx#1{{\partial #1\over\partial x}}
\def\pyy#1{{\partial #1\over\partial y}}
\def\pww#1{{\partial #1\over\partial w}}
\def\pss#1{{\partial #1\over\partial s}}
\def\prr#1{{\partial #1\over\partial r}}
\def\prhrh#1{{\partial #1\over\partial \rho}}
\def\pthth#1{{\partial #1\over\partial \theta}}
\def\pchch#1{{\partial #1\over\partial \chi}}
\def\ppsps#1{{\partial #1\over\partial \psi}}
\def\pzeze#1{{\partial #1\over\partial \zeta}}
\def\pphph#1{{\partial #1\over\partial \phi}}
\def\ptt#1{{\partial #1\over\partial t}}
\def\pVV#1{{\partial #1\over\partial V}}
\def\phh#1{{\partial #1\over\partial \theta}}
\def\pvhvh#1{{\partial #1\over\partial \vartheta}}
\def\pxixi#1{{\partial #1\over\partial \xi}}
\def\dtt#1{{d #1\over dt}}
\def\dss#1{{d #1\over ds}}
\def\drr#1{{d #1\over dr}}
\def\pprr#1{{\partial^2 #1\over\partial r^2}}
\def\pprhrh#1{{\partial^2 #1\over\partial \rho^2}}
\def\ppss#1{{\partial^2 #1\over\partial s^2}}
\def\ppxx#1{{\partial^2 #1\over\partial x^2}}
\def\ppxy#1{{\partial^2 #1\over\partial x\partial y}}
\def\ppxs#1{{\partial^2 #1\over\partial x\partial s}}
\def\ppys#1{{\partial^2 #1\over\partial y\partial s}}
\def\ppyy#1{{\partial^2 #1\over\partial y^2}}
\def\ppzz#1{{\partial^2 #1\over\partial z^2}}
\def\pptt#1{{\partial^2 #1\over\partial t^2}}
\def\ppVV#1{{\partial^2 #1\over\partial V^2}}
\def\ppphph#1{{\partial^2 #1\over\partial \phi^2}}
\def\ppthth#1{{\partial^2 #1\over\partial \theta^2}}
\def\pphh#1{{\partial^2 #1\over\partial \theta^2}}
\def\ppvhvh#1{{\partial^2 #1\over\partial \vartheta^2}}
\def\ppxixi#1{{\partial^2 #1\over\partial \xi^2}}
\def\ppzeze#1{{\partial^2 #1\over\partial \zeta^2}}
\def\pphze#1{{\partial^2 #1\over\partial\theta\partial\zeta}}
\def\ppz#1{\partial #1/\partial z}
\def\ppx#1{\partial #1/\partial x}
\def\ppy#1{\partial #1/\partial y}
\def\ppw#1{\partial #1/\partial w}
\def\ppr#1{\partial #1/\partial r}
\def\pprh#1{\partial #1/\partial \rho}
\def\pps#1{\partial #1/\partial s}
\def\ppt#1{\partial #1/\partial t}
\def\ppV#1{\partial #1/\partial V}
\def\pph#1{\partial #1/\partial \theta}
\def\ppvh#1{\partial #1/\partial \vartheta}
\def\ppxi#1{\partial #1/\partial \xi}

\def\ddt#1{d #1/dt}
\def\pppz#1{\partial^2 #1/\partial z^2}
\def\pppx#1{\partial^2 #1/\partial x^2}
\def\pppy#1{\partial^2 #1/\partial y^2}
\def\pppr#1{\partial^2 #1/\partial r^2}
\def\ppprh#1{\partial^2 #1/\partial \rho^2}
\def\ppps#1{\partial^2 #1/\partial s^2}
\def\pppt#1{\partial^2 #1/\partial t^2}
\def\pppV#1{\partial^2 #1/\partial V^2}
\def\ppph#1{\partial^2 #1/\partial \theta^2}
\def\pppvh#1{\partial^2 #1/\partial \vartheta^2}
\def\pppxi#1{\partial^2 #1/\partial \xi^2}
\def\dddt#1{d^2 #1/dt^2}

\def\grad{\nabla}
\def\cross{{\bf \times}}
\def\div{\grad\cdot}
\def\divp{\grad_\perp\cdot}
\def\divpl{\grad_\parallel\cdot}
\def\curl{\grad\cross}
\def\dpl{\grad_\parallel}
\def\ddpl{\grad_\parallel^2}
\def\dpp{\grad_\perp}
\def\ddpp{\grad_\perp^2}
\def\delsq{\grad^2}
\def\delamb{ \mathchar"0274\hskip -.665em\mathchar"0275 }
\let\delam=\delamb
\def\lapl{\grad^2}
\def\lapldef{\ddpp=(\pt^2/\pt x^2)+K^2(\pt^2/\pt y^2)}

\def\pwww#1{{\partial #1\over\partial \vec w}}
\def\pwwpl#1{{\partial #1\over\partial w_\parallel}}
 
\def\pvv#1#2{{\partial #2\over\partial v_{#1}}}
\def\ppv#1#2{{\partial #2/\partial v_{#1}}}
\def\pvvv#1{{\partial #1\over\partial \vec v}}
\def\pvvp#1#2{{\partial #2\over\partial v'_{#1}}}
\def\ppvp#1#2{{\partial #2/\partial v'_{#1}}}
\def\pvvvp#1{{\partial #1\over\partial \vec v'}}
\def\pvvpl#1{{\partial #1\over\partial v_\parallel}}
 
\def\xunit{\vec{\hat x}}
\def\yunit{\vec{\hat y}}
\def\zunit{\vec{\hat z}}
\def\sunit{\vec{\hat s}}
\def\bunit{\vec{b}}
\def\eunit{\vec{\hat e}}
\def\nunit{\vec{\hat n}}
\def\dt{\Delta t}
\def\becomes{\leftarrow}
\def\from{\leftarrow}
\def\to{\rightarrow}
\def\fromto{\leftrightarrow}
\def\implies{\,\,\,\Longrightarrow\,\,\,}
\def\dotdot{\!:\!}

\def\meters{\,{\rm m}}
\def\invm{\,{\rm m}^{-3}}
\def\invmeter{\,{\rm m}^{-1}}
\def\invsec{\,{\rm sec}^{-1}}
\def\cm{\,{\rm cm}}
\def\km{\,{\rm km}}
\def\invcc{\,{\rm cm}^{-3}}
\def\invcm{\,{\rm cm}^{-1}}
\def\invmm{\,{\rm mm}^{-1}}
\def\mm{\,{\rm mm}}
\def\Vcm{\,{\rm V/cm}}
\def\Acm{\,{\rm A/cm^2}}
\def\kA{\,{\rm kA}}
\def\MA{\,{\rm MA}}
\def\degk{\,{\rm K}}
\def\ergs{\,{\rm erg}}
\def\eV{\,{\rm eV}}
\def\keV{\,{\rm keV}}
\def\MeV{\,{\rm MeV}}
\def\GeV{\,{\rm GeV}}
\def\kG{\,{\rm kG}}
\def\tesla{\,{\rm T}}
\def\kW{\,{\rm kW}}
\def\MW{\,{\rm MW}}
\def\MWsqm{\,{\rm MW/m^2}}
\def\Wsqm{\,{\rm W/m^2}}
\def\radsec{\,{\rm rad/sec}}
\def\Hz{\,{\rm Hz}}
\def\kHz{\,{\rm kHz}}
\def\MHz{\,{\rm MHz}}
\def\mpersec{\,{\rm m}/{\rm sec}}
\def\msqsec{\,{\rm m^2}/{\rm sec}}
\def\cmsec{\,{\rm cm}/{\rm sec}}
\def\kmsec{\,{\rm km}/{\rm sec}}
\def\mmsec{\,{\rm m}^2/{\rm sec}}
\def\msqsec{\,{\rm m}^2/{\rm sec}}
\def\cmcmsec{\,{\rm cm}^2/{\rm sec}}
\def\ccpersec{\,{\rm cm}^3/{\rm sec}}
\def\minutes{\,{\rm min}}
\def\yr{\,{\rm yr}}
\def\hr{\,{\rm hr}}
\def\Bar{\,{\rm bar}}
\def\sec{\,{\rm sec}}
\def\msec{\,{\rm msec}}
\def\usec{\,\mu{\rm sec}}

\def\ee{\vec E}
\def\bb{\vec B}
\def\ff{\vec F}
\def\jj{\vec J}
\def\qq{\vec q}
\def\aa{\vec A}
\def\kk{\vec k}
\def\vv{\vec v}
\def\uu{\vec u}
\def\xx{\vec x}
\def\ww{\vec w}

\def\bdel{\vec b\cdot\grad}
\def\Bdel{\vec B\cdot\grad}
\def\Jdel{\vec J\cdot\grad}
\def\bdot{\vec B\cdot}
\def\Bdot{\vec B\cdot}
\def\kdot{\vec k\cdot}
\def\exb{\vec E\cross\vec B}
\def\jxb{\vec J\cross\vec B}
\def\uxb{\vec u\cross\vec B}
\def\vxb{\vec v\cross\vec B}
\def\wxb{\vec w\cross\vec B}
\def\ucxb{{\vec u\over c}\cross\vec B}
\def\vcxb{{\vec v\over c}\cross\vec B}
\def\wcxb{{\vec w\over c}\cross\vec B}
\def\jcxb{{\vec J\cross\vec B\over c}}

\def\vexb{\vec v_E}
\def\vpol{\vec v_p}
\def\upol{\vec u_p}
\def\vstar{\vec v_*}
\def\ustar{\vec u_*}
\def\Jstar{\vec J_*}
\def\Jpol{\vec J_p}
\def\vgradb{\vec v_{\grad B}}
\def\qpol{\vec q_p}
\def\qstar{\vec q_\wedge}
\def\qestar{\vec q_e{}_\wedge}
\def\qistar{\vec q_i{}_\wedge}
\def\pistar{\vec\Pi_*}
\def\vR{\vec v_R}
\def\vdl{\vec v\cdot\grad}
\def\vdel{\vec v\cdot\grad}
\def\vedl{\vexb\cdot\grad}
\def\udl{\vec u\cdot\grad}
\def\udel{\vec u\cdot\grad}
\def\uidl{\vec u_i\cdot\grad}
\def\uidel{\vec u_i\cdot\grad}
\def\wdel{\vec w\cdot\grad}
\def\dedt#1{d_E #1/dt}
\def\dett#1{{d_E #1\over dt}}
\def\jpp{J_\perp}
\def\jperp{\vec\jpp}
\def\qpp{q_\perp}
\def\qperp{\vec\qpp}
\def\upp{u_\perp}
\def\uperp{\vec\upp}
\def\wpl{w_\parallel}
\def\wpp{w_\perp}
\def\wperp{\vec\wpp}
\def\vpp{v_\perp}
\def\vperp{\vec\vpp}
\def\lnb{\log B}
 
\def\rms{_{rms}}
 
\def\Jpl{J_\parallel}
\def\jpl{J_\parallel}
\def\Jpp{J_\perp}
\def\jpp{J_\perp}
\def\Jperp{\vec\Jpp}
\def\Bperp{\vec B_\perp}
\def\Apl{A_\parallel}
\def\apl{A_\parallel}
\def\App{A_\perp}
\def\app{A_\perp}
\def\Aperp{\vec\App}
\def\Epl{E_\parallel}
\def\epl{E_\parallel}
\def\Epp{E_\perp}
\def\epp{E_\perp}
\def\Eperp{\vec\Epp}
\def\upl{u_\parallel}
\def\vpl{v_\parallel}
\def\Upl{U_\parallel}
\def\vor{\grad_\perp^2\phi}
\def\kpl{k_\parallel}
\def\kkpl{k_\parallel^2}
\def\kpp{k_\perp}
\def\kperp{\vec\kpp}
\def\kkpp{k_\perp^2}
\def\xpl{{x_\parallel}}
\def\xpp{x_\perp}
\def\DD{\Delta_D}
\def\Dpl{D_\parallel}
\def\Dpp{\Delta_\perp}
\def\Depl{D_e{}_\parallel}
\def\Dipl{D_i{}_\parallel}
\def\Rpl{R_\parallel}
\def\qpl{q_\parallel}
\def\qepl{q_e{}_\parallel}
\def\qipl{q_i{}_\parallel}
\def\Pipl{\Pi_\parallel}
\def\qeperp{\vec q_e{}_\perp}
\def\qiperp{\vec q_i{}_\perp}
\def\mupl{\mu_\parallel}
\def\mupp{\mu_\perp}
\def\nuei{\nu_{ei}}
\def\nuee{\nu_{ee}}
\def\nuii{\nu_{ii}}
\def\wpe{\omega_{pe}}
\def\wpi{\omega_{pi}}
\def\nudamp{\nu_d}
\def\zeff{Z_{\!e\!f\!f}}
\def\lmfp{\lambda_{\!m\!f\!p}}
\def\ws{{\omega_*}}
\def\wsi{{\omega_{*i}}}
\def\wn{\omega_n}
\def\wt{\omega_t}
\def\wi{\omega_i}
\def\wT{\omega_T}
\def\wp{\omega_p}
\def\wc{{\omega_c}}
\def\kappacv{{\cal K}}
\def\wcv{{\omega_B}}
\def\etai{\eta_i}
\def\taui{\tau_i}
\def\rs{\rho_s}
\def\ld{\lambda_D}
\def\Lpl{L_\parallel}
\def\Lpp{L_\perp}
\def\lcorpl{\lambda_\parallel}
\def\lcorpp{\lambda_\perp}
\def\lcorx{\lambda_x}
\def\lcory{\lambda_y}
\def\rch{\rho_{ch}}
\def\npl{\eta_\parallel}
\def\etapl{\eta_\parallel}
\def\ald{a_L}
\def\alde{a_{Le}}
\def\aldi{a_{Li}}
\def\npp{\eta_\perp}
\def\etapp{\eta_\perp}
\def\kappapl{\kappa_\parallel}
\def\dprime{\Delta'}
\def\sk{{}_{\vec k}}
\def\sky{{}_{k_y}}
\def\gk{\gamma_k}
\def\vk{\vfl_k}
\def\nk{\nfl_k}
\def\tk{\tfl_k}
\def\dk{\Delta k}
\def\gd{\gamma_0}
\def\mwn{\Delta_n}
\def\mwh{\Delta_h}
\def\gamT{\Gamma_T}
\def\gamn{\Gamma_n}
\def\gamt{\Gamma_t}
\def\gami{\Gamma_i}
\def\gamc{\Gamma_c}
\def\gamk{\Gamma_k}
\def\gams{\Gamma_s}
\def\gaml{\Gamma_l}
\def\gamr{\Gamma_r}
 
\def\ptb{\widetilde}
\def\psifl{\widetilde\psi}
\def\phifl{\widetilde\phi}
\def\ffl{\widetilde f}
\def\fe{f_e}
\def\fefl{\widetilde f_e}
\def\fifl{\widetilde f_i}
\def\nfl{\widetilde n}
\def\hfl{\widetilde h}
\def\tfl{\widetilde T}
\def\nefl{\widetilde n_e}
\def\nifl{\widetilde n_i}
\def\tefl{\widetilde T_e}
\def\tifl{\widetilde T_i}
\def\pfl{\widetilde p}
\def\pefl{\widetilde p_e}
\def\pifl{\widetilde p_i}
\def\hefl{\widetilde h_e}
\def\vx{\widetilde v_x}
\def\vfl{\widetilde v}
\def\vefl{\widetilde \vexb}
\def\vxfl{\widetilde v_x}
\def\vyfl{\widetilde v_y}
\def\vrfl{\widetilde v_r}
\def\vppfl{\widetilde v_\perp}
\def\vflpp{\widetilde v_\perp}
\def\vplfl{\widetilde \vpl}
\def\Bfl{\widetilde \vec B}
\def\Bflpp{\widetilde B_\perp}
\def\Aplfl{\widetilde A_\parallel}
\def\Appfl{\widetilde A_\perp}
\def\Aperpfl{\widetilde {\vec A}_\perp}
\def\ufl{\widetilde u_\parallel}
\def\vorfl{\grad_\perp^2\phifl}
\def\jfl{\widetilde J_\parallel}
\def\qfl{\widetilde q_\parallel}
\def\qefl{\widetilde q_e{}_\parallel}
\def\qifl{\widetilde q_i{}_\parallel}
\def\jppfl{\widetilde J_\perp}
\def\jperpfl{\widetilde {\vec J}_\perp}
\def\Afl{\ptb A_\parallel}
\def\Jfl{\ptb J_\parallel}
\def\efl{\widetilde E_\parallel}
\def\Efl{\widetilde E_\parallel}
\def\Eppfl{\widetilde E_\perp}
\def\Eperpfl{\widetilde {\vec E}_\perp}
\def\etafl{\widetilde\eta}
\def\isatfl{\widetilde I_{{\rm sat}}}
\def\phiflfl{\widetilde\phi_{{\rm fl}}}
 
\def\teplfl{\widetilde T_e{}_\parallel}
\def\teppfl{\widetilde T_e{}_\perp}
\def\qeplfl{\widetilde q_e{}_\parallel}
\def\qeppfl{\widetilde q_e{}_\perp}
\def\tiplfl{\widetilde T_i{}_\parallel}
\def\tippfl{\widetilde T_i{}_\perp}
\def\qiplfl{\widetilde q_i{}_\parallel}
\def\qippfl{\widetilde q_i{}_\perp}

\def\tepl{ T_e{}_\parallel}
\def\tepp{ T_e{}_\perp}
\def\qepl{ q_e{}_\parallel}
\def\qepp{ q_e{}_\perp}
\def\tipl{ T_i{}_\parallel}
\def\tipp{ T_i{}_\perp}
\def\qipl{ q_i{}_\parallel}
\def\qipp{ q_i{}_\perp}

\def\peplfl{\widetilde p_e{}_\parallel}
\def\peppfl{\widetilde p_e{}_\perp}
\def\piplfl{\widetilde p_i{}_\parallel}
\def\pippfl{\widetilde p_i{}_\perp}

\def\pepl{ p_e{}_\parallel}
\def\pepp{ p_e{}_\perp}
\def\pipl{ p_i{}_\parallel}
\def\pipp{ p_i{}_\perp}


\def\phinn{ {e\phifl\over T} }
\def\nnn{ {\nfl\over n} }
\def\tnn{ {\tfl\over T} }
\def\unn{ {\ufl\over c_s} }
\def\vornn{ \rho_s^2\ddpp\phinn }
\def\jnn{ {\jfl\over ne} }
\def\qnn{ {\qfl\over nT} }
\def\psinn{ {\psifl\over B\rho_s} }

\def\ahat{\hat\alpha}
\def\ehat{\hat\eta}
\def\khat{\hat\kappa}
\def\shat{\hat s}
\def\bhat{\hat\beta}
\def\muhat{\hat\mu}
\def\epss{\hat\epsilon}
\def\bigpoint#1{
    \par\bigskip
    {\baselineskip=\normalbaselineskip
    \parindent=0 pt
    {\hfill\vbox{ #1  }\hfill}}
    \par\bigskip
    }
 
\def\jfm#1{{\it J. Fluid. Mech.} {\secfnt #1}}
\def\jgr#1{{\it J. Geophys. Res.} {\secfnt #1}}
\def\prl#1{{\it Phys. Rev. Lett.} {\secfnt #1}}
\def\physletta#1{{\it Phys. Lett. A} {\secfnt #1}}
\def\physlettb#1{{\it Phys. Lett. B} {\secfnt #1}}
\def\pf#1{{\it Phys. Fluids} {\secfnt #1}}
\def\pfa#1{{\it Phys. Fluids A} {\secfnt #1}}
\def\pfb#1{{\it Phys. Fluids B} {\secfnt #1}}
\def\physp#1{{\it Phys. Plasmas} {\secfnt #1}}
\def\nf#1{{\it Nucl. Fusion} {\secfnt #1}}
\def\njp#1{{\it New J. Phys.} {\secfnt #1}}
\def\cpp#1{{\it Contrib. Plasma Phys.} {\secfnt #1}}
\def\ppcf#1{{\it Plasma Phys. Contr. Fusion} {\secfnt #1}}
\def\plasphys#1{{\it Plasma Phys.} {\secfnt #1}}
\def\revpp#1{{\it Rev. Plasma Phys.} {\secfnt #1}}
\def\iaea#1#2{in {\it Plasma Physics and Controlled Nuclear Fusion
    Research #1}, Proceedings of the #2th International Conference}
\def\EPS#1#2#3{in {\it Proceedings of the
{#1}th European Conference on Controlled Fusion and Plasma Physics,
{#2}, {#3}} (European Physical Society, {#2}, {#3})}
\def\jcp#1{{\it J. Comput. Phys.} {\secfnt #1}}
\def\jetp#1{{\it Sov. Phys. JETP} {\secfnt #1}}
\def\sovjpp#1{{\it Sov. J. Plasma Phys.} {\secfnt #1}}
\def\jnm#1{{\it J. Nucl. Mat.} {\secfnt #1}}
\def\rsi#1{{\it Rev. Sci. Inst.} {\secfnt #1}}
\def\adv#1{{\it Adv. Phys.} {\secfnt #1}}
\def\apjl#1{{\it Astrophys. J. Lett.} {\secfnt #1}}
\def\apj#1{{\it Astrophys. J.} {\secfnt #1}}
\def\astrap#1{{\it Astron. Astrophys.} {\secfnt #1}}
\def\mnras#1{{\it MNRAS} {\secfnt #1}}
\def\vol#1{\ {\secfnt #1}}

\def\zdot{\dot z}
\def\Rdot{\vec{\dot R}}
\def\thetadot{\dot\vartheta}
\def\zpdot{\vec{\dot Z_p}}
\def\pzzp#1{{\pt #1\over\pt\vec Z_p}}
\def\pvpvp#1{{\pt #1\over\pt\varphi}}
\def\ppzpz#1{{\pt #1\over\pt p_z}}
\def\diffa{{\vec d_a}}
\def\diffb{{\vec d_b}}
\def\subR{{}_{\vec R}}

\def\Epsilon{{\cal E}}
\def\Epslash{{\cal E}\hskip -0.2 cm / \hskip 0.05 cm}
\def\epslash{\epsilon\hskip -0.15 cm / \hskip 0.001 cm}

\def\sumsp{\sum_{{\rm sp}}}
\def\sumions{\sum_{{\rm ions}}}
\def\dL{d\Lambda}
\def\dW{d{\cal W}}
\def\dV{d{\cal V}}
\def\scriptl{{\cal L}}
\def\scripte{{\cal E}}
\def\scriptp{{\cal P}}
\def\scriptq{{\cal Q}}
\def\pmumu#1{{\pt #1\over\pt\mu}}
\def\magnet{{\cal M}}

\def\Astar{\vec A^*}
\def\Aphi{A^*_\varphi}
\def\bphi{b_\varphi}
\def\Pphi{P_\varphi}
\def\Bstar{\vec B^*}
\def\Bpl{B_\parallel^*}
\def\Bpp{B_\perp}
\def\chiv{\chi'}
\def\Bfl{\ptb B}

\def\Tpl{T_\parallel}
\def\Tpp{T_\perp}

\def\vefl{\ptb v}
\def\vexfl{\ptb v^V}
\def\veyfl{\ptb v^\xi}
\def\velyfl{\ptb v_\xi}
\def\velpfl{\ptb v_\vartheta}
\def\velp{v_\vartheta}

\def\bfl{\ptb b}
\def\bxfl{\ptb b^V}
\def\byfl{\ptb b^\xi}
\def\blyfl{\ptb b_\xi}
\def\blpfl{\ptb b_\varphi}

\section{Introduction}

Gyrokinetic theory is a well founded formalism by which particle motion
is treated in terms of drifts of particle gyrocenters rather than the
combination of gyromotion and drifts of the particles.  Particle motion
in a magnetic field is set up with a drift kinetic Lagrangian assuming
arbitrarily large magnetic field scale length 
\cite{Littlejohn81,Littlejohn83,WhiteChance84}, and then
Lie transform techniques assuming small product of gyroradius and field
amplitude are applied to obtain a Lagrangian still independent
of gyrophase angle but valid for a gyroradius, while still small compared to
background scale lengths, of arbitrary order with respect to the scale
of E-cross-B eddies \cite{CaryLittlejohn83,Dubin83,Hahm88,Hahm88a}.  
Alternatively, the field variable amplitude may be left arbitrary when
the small parameter for expansion is the local ratio between the
gyroradius and the scale of potential variation \cite{Brizard95,Miyato09}.  
Initially in this formulation, a back transformation 
was used to obtain the self consistent field polarisation equation for
the electrostatic potential \cite{Dubin83,Hahm88}.
Taken together the gyrokinetic Lie transform and this back transform are
an application of push forward and pull back transforms in differential
geometry \cite{CaryLittlejohn83,Qin04,Brizard07}.
More recently the entire Lagrangian is set
up as the integral of a Lagrangian density over the phase space, with
the polarisation equation obtained as the Euler-Lagrange equation by
varying the electrostatic potential in the field Lagrangian
\cite{Sugama00,Brizard00}.  These two approaches were shown to be
equivalent in the recent review by Brizard and Hahm \cite{Brizard07}.
Moreover, in an analysis of gyrokinetic transformation of the general
Landau collision operator, the method of Lie transforming the Lagrangian
and deriving the Euler-Lagrange equations was shown to be equivalent to
the Poisson bracket transform of the Vlasov or Boltzmann kinetic
equation directly, with the latter method able to treat collisional
dissipation but with the Lie transform useful in deciding which
coordinate map to use \cite{Brizard04}.
Following demonstrative gyrokinetic 
simulation of the internal kink mode \cite{Naitou95},
gyrokinetic theory was explicitly linked back to MHD for long wavelength
electromagnetic oscillations and instabilities
\cite{Qin99,Qin00,Lee01,Lee03}, 
Using the large amplitude/long wavelength
form of Ref.\ \cite{Miyato09} the correspondence to nonlinear reduced
magnetohydrodynamics (MHD) and with the long wavelength limit of the
small amplitude version was shown at the level of the Lagrangians.
The theory is now a fully self consistent
Lagrangian field theory.  As it does not depend on assumptions
concerning the form of the distribution function, it is also necessarily
a total-f formulation.

Energy conservation has been well known since the development of
the Lie transform version of the theory as cited above.  Momentum has
received less attention despite the general demonstration of
conservation via the Noether theorem \cite{Sugama00}.
However, since the rise of gyrokinetic
treatments of neoclassical flows \cite{WWang01,WWang04}, discussion of
the Coriolis drift and turbulent equipartition effects
\cite{Peeters07,Hahm07,Hahm08,Peeters09}\ has emerged during observation
of the tokamak momentum pinch \cite{deGrassie09}.  
The turbulent equipartition scenario as an indirect effect involving
exchange channels more than actual drive effects
especially underscores the role of strict conservation in a complicated,
nonlinear physical situation.
Momentum conservation was also
demonstrated for evolution of axisymmetric flows and currents toward
equilibrium in the context of total-f  
electromagnetic theory and computation under edge
conditions \cite{pet09}.  

Gyrokinetic field theory was not necessary to build the
original gyrokinetic computational models whose self consistent
equation for the field potential was referred to as the gyrokinetic
Poisson equation \cite{Lee83,Lee87,Parker93}.  Despite recovery of these
equations by the Lie transform and field theory methods cited above,
the usefulness of the gyrokinetic Poisson equation for determining the
electric field has been criticised \cite{Catto08}.  Within the field
theory, however, the Euler-Lagrange equation for the electrostatic
potential is indeed one and the same with this gyrokinetic Poisson
equation \cite{Sugama00}.  Obtaining the equation via different methods
outside of the field theory would appear to break the inherent
consistency unless the method were found to be equivalent to use of the
field theory, as the earlier versions in fact are.

We therefore address the general question of momentum conservation
within the energetic consistency afforded by the field theory version of
gyrokinetics.  The results of any particular Lie transform are assumed
to be given, with the only stipulation being the general form of a
Lagrangian in which all dependence on time dependent field variables is
transformed into the Hamiltonian (\ie, the ``interaction Lagrangian'' as
described in basic texts \cite{LandauLifshitz}).  Both energy and
momentum are considered, with the role of symmetries as paramount.  In a
tokamak under low frequency conditions, the conserved momentum is
toroidal momentum; the other components of the momentum vector use the
spatially dependent background magnetic field as an anchor.  We leave
the issue of a conserved momentum vector, a tensor transport flux, and
the anchor to future work and concentrate on toroidal momentum.  The
result is that momentum transport and conservation yields familiar
content and transport fluxes, generally and even in the specific cases
of conventional models.  The route back to MHD is explicitly shown via
appropriate choice of the Lagrangian/Hamiltonian.  The main results are
valid for any ordering scheme which might be used since their
demonstration does not depend on ordering but uses general functional
form of the Hamiltonian on the field variables and a field term
describing shear-Alfv\'en disturbance magnetic energy.

Following sections of the paper describe the field theory model in
general with emphasis on energetic consistency, 
recover global conservation laws via the Noether theorem,
address the conservation of generalised vorticity, and then address
toroidal momentum and energy.  A 4-dimensional antisymmetric bracket
form of the gyrokinetic equation is derived (the 5th and 6th dimensions
do not enter due to the conservation of generalised magnetic moment and
the lack of dependence by any dependent variable on gyrophase angle),
which not only greatly facilitates the mathematics but is also suitable
for computations \cite{pet09}.
Two comments on orderings are given, both
how na\"\i ve ordering schemes can violate energetic consistency, and on
a result showing that momentum evolution via small
fluctuations occur in two
places via terms of the same order (\ie, orderings that occur there
enter order by order in the same term multiplied by factors of order
unity).  The general momentum and energy transport equations that are
derived are independent of any ordering.  The route back to MHD is shown
using what can be called an ``MHD Hamiltonian'' is shown explicitly for
momentum and a mean field fluctuation model within that is given.
The salient mathematics regards application of functional derivatives,
for which the background and main operations are given in Appendix A.
Then, Appendix B treats conventional models and their equivalence to
gyrokinetic field theory versions in each case via appropriate choice of
the Lagrangian.  The results on momentum depend on a cancellation
mandated by the transport equation for generalised vorticity;
Appendix C gives the version of this by considering the torque due to a
charge source under quasineutral conditions.

\section{Description of the gyrokinetic field theory model}

The general gyrokinetic theory follows from Lie transforms applied to
the extended phase space Lagrangian for the particle gyrocenter motion
\cite{CaryLittlejohn83,Dubin83,Hahm88}.  
A phase space kernel locating the particles
transforms 
the Lagrangian to a Lagrangian density \cite{Sugama00}.  In the
electrostatic version the phase space integral over this density is
the entire action.  In the case of an electromagnetic model the magnetic
energy is added as a free field contribution as in general
electrodynamics (for background see the text by Landau and Lifshitz
\cite{LandauLifshitz}).
The electric field energy contribution is neglected since in a
magnetised plasma the ExB kinetic energy of the particle drifts is much
larger; this is the same statement as quasineutrality since a zero space
charge density is the natural result \cite{Sugama00}.  Variation of this
action with respect to the gyrocenter coordinates gives the
Euler-Lagrange equations for the particles.  Liouville's theorem is used
to convert these into an equation for a distribution function, which
serves as the gyrokinetic Vlasov equation.  Variation of the action with
respect to the field potentials gives the polarisation and induction
equations giving the self consistent response of the electrostatic
potential and parallel magnetic potential, respectively
\cite{Sugama00,Brizard00}.  

Gyrokinetic theory was recently given in
terms of the field theory for transonic ExB flows \cite{Miyato09}, 
readdressing the large amplitude (strong ExB flow) version of
Ref.\ \cite{Brizard95}.  Following Ref.\ \cite{Hahm96}, the ExB flow
resulting from the appearance of the large-scale potential 
in the lowest-order Euler-Lagrange equations
was used in the coordinate
transformations, rather than a background flow given by a fluid
analysis.  In our case, however, the same potential was used
for all flow dynamics without splitting the fluctuations from the
background; the dependent field variable was used for both purposes.
The choice of Lie transform was then changed to move all effects from
the flow potential into the Hamiltonian.  The model was shown to recover
the conventional ones \cite{Hahm88,Hahm88a},
including nonlinear reduced MHD
\cite{Strauss76,Strauss77}, for weaker flow amplitude and larger scale.

The particle Lagrangian itself starts with the form from electrodynamics
and is Lie transformed into a low-frequency form in an expansion which
formally uses the amplitude of the drifts as a small parameter; this can
be either the fluctuation amplitude as in Refs.\ \cite{Dubin83,Hahm88}
or the gyroradius compared to the dynamical scale length as in
Refs.\ \cite{Brizard95,Miyato09}.  

The Lie transform method is rather general, and various choices exist,
but for present purposes it is useful to know that the choice can always
be made to arrange the particle Lagrangian, $L_p$, so that the
symplectic part (the vector of coefficients, $p_i$,
of particle coordinate time
derivatives, $\dot q^i$, in the
representation $L_p=p_i\dot q^i-H$)
depends on background geometry only, and all time dependent
field effects appear in the Hamiltonian, $H$, only.  Since the latter
serves as the time component of the underlying fundamental one-form,
this casts the Euler-Lagrange equations in a form where partial time
derivatives on dynamical field variables are absent.  Resulting
geometric quantities, including volume elements and Jacobians,
are strictly static.
This was the Lie-transform strategy used in Ref.\ \cite{Miyato09}, 
emphasising the general 
desirability of arranging $L_p$ with all time dependence in $H$ not only
for computations but also in proving correspondence to other forms such
as nonlinear reduced MHD.
For the case of tokamak geometry, this has the
added benefit of isolating all toroidal angle dependence into $H$ as
well, which will be seen to facilitate the proof of results involving
toroidal momentum conservation.

We therefore assume a particle Lagrangian which for any degree of
expansion has been Lie transformed into the following form
\begin{equation}
L_p = \LP{e\over c}\vec A+p_z\bunit\RP\cdot\dot\vec R
	+ {mc\over e}\mu\thetadot - H
\label{eqlagrangian}
\end{equation}
where $\{\vec Z_p\}=\{\vec R,p_z,\mu\}$ 
are the particle coordinates (gyrocenter
position, parallel canonical momentum, 
gyration magnetic moment), $\vartheta$ is the
ignorable sixth coordinate giving the gyrophase
angle, $e$ and $m$ are the species charge and mass, 
$\vec A$ and $\bunit$ are the potential and
unit vector for $\vec B=\curl\vec A$ the background magnetic field, and
$H$ is the Hamiltonian.  The canonical parallel momentum version is
used, so that the time dependent parallel
magnetic potential $\Apl$ appears only in $H$.  All flow dynamics due to
the time dependent electrostatic potential $\phi$ appear also only in
$H$.  Here and below, the parallel subscript denotes the component
locally parallel to $\vec B$.

The Hamiltonian depends on both
field potentials $\phi$ and $\Apl$, evaluated at the gyrocenter
positions
via, e.g., $\phi(\vec R)$, as well as the gyrocenter phase space
coordinates, 
\begin{equation}
H = H(\vec R,p_z,\mu,\phi,\Apl)
\label{eqhamiltonian}
\end{equation}
The parallel velocity $U$ is not used explicitly as a coordinate
but can be defined as
a derivative of $H$,
\begin{equation}
U\equiv \ppzpz{H}
\end{equation}
It is important in all the derivations to note that $U$ has
spatial and time derivatives through its dependence on $\Apl$.

The dependence of $H$ upon the fields $\phi,\Apl$ involves differential
operators such as spatial derivatives
which commute generally with variations or
spatial or time derivatives,
or the gyroaveraging operator $J_0$ which depends on $\mu$ and $B$ as
well as spatial derivatives.  One has to know whether these operators
commute with derivatives.  With all time and toroidal angle dependence
transformed into $H$ we ensure to be working with a representation in
which this is true generally for differentiation with respect to 
either time or toroidal angle although not for all components of the
gradient operator. 

Formally, $J_0$ has the form in
wavenumber space of multiplication of Fourier coefficients by
the zeroth Bessel function
$J_0(\kpp\rho_L)$, where $\rho_L$ is the particle gyroradius given by 
$\rho_L=\vpp/\abs{eB/mc}$, with gyrofrequency
$\abs{eB/mc}$, or in terms of the coordinates by
$\rho_L^2=2\mu B/[m(eB/mc)^2]$.  The perp subscript denotes the component
in the plane locally perpendicular to $\vec B$.  Hence $J_0$ is time
symmetric in any geometry but toroidal angle symmetric only in tokamak
geometry.  The $J_0$ operator may be cast as a series of perpendicular
Laplacians, $\ddpp$, so in the local transport equations to be derived
below it is sufficient to consider $H$ with arbitrary dependence on
field amplitude and the field gradient and Laplacian.  This becomes
necessary when considering the role of functional derivatives in the
theory.

The particle equations of motion are
found from the Euler-Lagrange equations resulting from $L_p$.
The drift motion,
\begin{equation}
\Bpl\dtt{\vec R} = 
\grad H\cdot{c\over e}{\vec F\over B} + U\Bstar 
\qqquad
	\Bpl\dtt{p_z} = -\Bstar\cdot\grad H
\label{eqdriftmotion}
\end{equation}
separates naturally from the gyromotion,
\begin{equation}
\dtt{\mu} = 0 \qqquad \dtt{\vartheta} = {e\over mc}\pmumu{H}
\label{eqgyromotion}
\end{equation}
where some standard definitions are
\begin{equation}
\Astar=\vec A + p_z {c\over e}\bunit \qqquad  
\Bstar=\curl\Astar \qqquad \Bpl=\bunit\cdot\Bstar
\end{equation}
Drift tensor notation is used, with 
\begin{equation}
\vec F = \grad\vec A-(\grad\vec A)^T
\end{equation}
where superscript $T$ denotes the transpose.
It follows that
\begin{equation}
\vec F = \eps\cdot\vec B \qqquad \curl\bunit = -\div{\vec F\over B}
\qqquad\Bstar = \vec B - p_z\div{c\over e}{\vec F\over B}
\end{equation}
where $\eps$ is the rank-three Levi-Civita pseudotensor.
Phase space volume conservation is expressed by
\begin{equation}
\pzzp{}\cdot(\sqrt{g}\Bpl\zpdot) = 0
\label{eqphasespace}
\end{equation}
where $\sqrt{g}$ is the determinant of covariant components of the
coordinate metric.  
With $\div\Bstar=0$, Eq.\ (\ref{eqphasespace}) implies
\begin{equation}
\div {c\over e}{\vec F\over B} + {\pt\Bstar\over\pt p_z} = 0
\end{equation}
Guaranteed by the definition of $\Bstar$,
this also determines that the quantity $\Bpl$ serves as the volume
element of the velocity space, where the volume element of the entire
phase space is $\sqrt{g}\Bpl$.

This particle Lagrangian is converted to a system one by placing the
particles in phase space via the kernel $G(\vec Z_p,\vec Z)$ with 
\begin{equation}
\vec Z\to\{\vec x,z,w\} \qqquad \vec Z_p\to\{\vec R,p_z,\mu\}
\end{equation}
giving the correspondence between phase space coordinates $\vec Z$
and gyrocenter coordinates $\vec Z_p$, respectively.  Due to some notational
difficulties with the rest of this work, however, we will dispense with
this distinction between $\vec Z$ and $\vec Z_p$, leaving the role of
$G$ to be understood once we already have the particle equations of
motion in Eqs.\ (\ref{eqdriftmotion},\ref{eqgyromotion}).

The phase space
integral is denoted $\int\dL$.  The integration domain $\dL$ 
is given as a combination of the velocity space and configuration space
domains.  These are given respectively by
\begin{equation}
\dL = \dV\otimes\dW \qqquad \dV = \sqrt{g}\,dx^1\,dx^2\,dx^3 \qqquad 
\dW = 2\pi m^{-2}\, dp_z\,d\mu\,\Bpl
\label{eqdwelement}
\end{equation}
where $\sqrt{g}$ is the determinant of covariant components of the
coordinate metric, and
noting that the form of $\dW$ is determined by phase space conservation
(the $\Bpl$ factor) and normalisation (the $2\pi m^{-2}$ factor).
The Lagrangian for the entire particles/field system is then
\begin{equation}
L=\sumsp\int\dL\, f\,L_p - \int\dV\,{\Bpp^2\over 8\pi}
\label{eqaction}
\end{equation}
where the sum is over species.  The electrodynamic
field term $(E^2-B^2)/8\pi$ reduces
to $-\Bpp^2/8\pi$ as the assumption of
quasineutrality eliminates $E^2/8\pi$ in favour of the 
ExB kinetic energy of the particle drifts,
and the assumptions of 
low frequency $\omega\ll\kpp v_A$ 
and low plasma beta $\beta=8\pi p/B^2\ll 1$
restrict magnetic
variation to the parallel magnetic potential $\Apl$.  A simplified
version usable for the present purposes is
\begin{equation}
\Bpp^2 = \abs{\dpp\Apl}^2
\label{eqbpp}
\end{equation}
defined with the perpendicular spatial derivatives.  Hence $\Bpp^2/8\pi$
is identified as the energy in shear Alfv\'en magnetic disturbances
perpendicular to the equilibrium magnetic field.
More general forms are possible (cf.\ Appendix A) but the field term is
always quadratic in $\Apl$.  Under conditions of magnetic
compressibility, $\vec A_\perp$ enters as well (cf.\ Sec.\ III C of
Ref.\ \cite{Brizard07}), but the structure of the theory remains as
presented herein.

The gyrokinetic Vlasov equation is found via
variation of the $\vec Z_p$ components according to characteristic
methods \cite{Sugama00}, or equivalently by application of Liouville's
theorem to the particle motion in Eqs.\
(\ref{eqdriftmotion},\ref{eqgyromotion}), yielding
\begin{equation}
\Bpl\ptt{f} + \grad H\cdot{c\over e}{\vec F\over B}\cdot\grad f
	+ \Bstar\cdot\LP\ppzpz{H}\grad f-\ppzpz{f}\grad H\RP =0
\label{eqgkf}
\end{equation}
Derivatives with respect to $\mu$ or $\vartheta$ do not appear, 
because $f$ and $H$ are independent of $\vartheta$,
and $\mu$ is conserved in the gyromotion.  
Once we have the gyrokinetic equation in this form, the
distinction between phase space and gyrocenter
coordinates may be left implicitly understood, since we no longer
consider particles or gyrocenters as discrete entities.

The term ``drifts'' refers to the drift motion described by
Eq.\ (\ref{eqdriftmotion}), especially the spatial part $\Rdot$.  The
part resulting from the field dependent variables is entirely contained
in $H$.  Hence when we refer to the treatment of drifts we mean the
construction of $H$ and in particular drifts to a certain order means
the contributions to $H$ due to an expansion up to that order.  The
results we will obtain do not depend on the form of ordering (just on
the functional form of the dependence of $H$ upon $\phi$ and $\Apl$), 
but at certain points we will need to refer to the result in terms of a
certain ordering. 

The equations for the fields are determined by functional derivatives
(cf.\ Appendix A and the background references cited there).
The self consistent polarisation equation 
(also called gyrokinetic Poisson equation)
is given by the Euler-Lagrange equation for $\phi$ from this same
Lagrangian.  It is found by varying the
Lagrangian with respect to $\phi$, yielding an integral over $\dV$ of
$\delta\phi(\vec x)$ times a coefficient, which is required to vanish.
It is the same statement as requiring the functional derivative of $L$
with respect to $\phi$ to vanish.  This produces 
\begin{equation}
\sumsp{\delta fH\over\delta\phi} = 0
\label{eqpolarisation}
\end{equation}
The species-summed functional derivative of $fH$ vanishes alone 
because $\phi$ appears only in $H$.  The functional derivative implies
velocity space integration because it is defined with respect to the
space covered by $\dV$.
The functional derivative combination yields
the gyrokinetic charge density for the particular $H$ used (the
assumption of quasineutrality sets this to zero).

The self consistent induction equation 
(also called gyrokinetic Amp\`ere equation)
is obtained by variation of
the field potential $\Apl$.  
It is the same statement as requiring the functional derivative of $L$
with respect to $\Apl$ to vanish.  This produces 
\begin{equation}
\sumsp{\delta fH\over\delta\Apl} = {1\over 4\pi}\ddpp\Apl 
\label{eqinduction}
\end{equation}
with the field term appearing on the right side arising from the field
term $-\Bpp^2/8\pi$ in $L$.  The functional derivative combination yields
the gyrokinetic current (times $-1/c$) for the particular $H$ used.  

\subsection{Antisymmetric Bracket Form of the Gyrokinetic Equation}

It has been found previously that maximal symmetry in the representation
of the gyrokinetic equation (Eq.\ \ref{eqgkf}) is helpful to the
understanding of the conservation laws \cite{pet09}.  We observe that
\begin{equation}
\ppzpz{\Astar} = {c\over e}\bunit
\qquad\hbox{hence}\qquad
\ppzpz{}\epsilon\cdot\Astar = {c\over e}{\vec F\over B}
\end{equation}
If we define
\begin{equation}
\vec G = \epsilon\cdot\Astar
\end{equation}
we may recast Eq.\ (\ref{eqgkf}) as
\begin{equation}
\Bpl\ptt{f} + \grad H\cdot\ppzpz{\vec G}\cdot\grad f
	+ (-\div\vec G)\cdot\LP\ppzpz{H}\grad f-\ppzpz{f}\grad H\RP =0
\end{equation}
This has the structure of one 3-bracket of indices $\{abz\}$ 
\begin{equation}
[H,G^{ab},f]_{azb} = \ppzpz{G^{ab}}[H,f]_{ab} + 
(\grad_aG^{ab})[H,f]_{bz} + (\grad_bG^{ab})[H,f]_{za}
\end{equation}
for each
pair of spatial coordinates $\{ab\}$ with index $z$ denoting the $p_z$
coordinate. 
This can also be written as
\begin{equation}
[H,G^{ab},f]_{azb} = \epsilon^{abc}\LP\ppzpz{A^*_c}[H,f]_{ab} + 
(\grad_aA^*_c)[H,f]_{bz} + (\grad_bA^*_c)[H,f]_{za}\RP
\end{equation}
where on the right side the Einstein summation convention is used for
repeated (up/down) indices.
In each case the 2-bracket form is
\begin{equation}
[H,f]_{ab} = H_{,a}\,f_{,b} - H_{,b}\,f_{,a}
\end{equation}
with the comma denoting differentiation with respect to the coordinate
whose index is given by the subscript.
Since there is no $p_z$-component
of $\Astar$ we may add 3 more fictitious 3-brackets, one for each pair
of spatial indices and $p_z$ with $A^*_z$.  The entire combination becomes
\begin{equation}
\ptt{f} + \Epsilon^{abcd}H_{,a}f_{,b}A^*_{c,d} = 0
\label{eqgk4}
\end{equation}
where $\Epsilon$ is the rank-four Levi-Civita pseudotensor in the
4-space covered by $\dV\otimes dp_z$.  The components of $\epsilon^{abc}$
are $1/\sqrt{g}$ times $\pm 1$ or $0$ depending on the permutation of
spatial indices $\{abc\}$.  The components of $\Epsilon^{abcd}$
are $1/\sqrt{g}\Bpl$ times $\pm 1$ or $0$ depending on the permutation of
indices $\{abcd\}$ in the 4-space domain.  The 3-space order is
$\{123\}$ for $dx^1\,dx^2\,dx^3$ and hence the 4-space order is
$\{123z\}$ for $dx^1\,dx^2\,dx^3\,dp_z$.  Positive, negative, and zero
permutations of these give the other components.

It was previously observed that axisymmetric momentum conservation
follows directly from this form of the gyrokinetic equation, simply due
to symmetries in the indices \cite{pet09}.  In this work this
antisymmetric bracket form will be used to facilitate proof of the
conservation laws for energy and toroidal momentum for general
dependence of the Hamiltonian upon the dynamical fields.

\section{Global Conservation Laws}

The conserved energy is found from the total action
$\int dt\,L(\vec Z_p,\phi,\Apl,t)$ via Noether's theorem, applying
small variations to the time component \cite{LandauLifshitz}.  
In the gyrokinetic case this has been done before, both from the
discrete particle characteristics point of view \cite{Sugama00}, and
from the continuum/field representation using constrained variations
\cite{Brizard00}.  Since $L$
is first order in all the time derivatives this becomes the combination
of all the $p_i\dot q^i$ terms less the Lagrangian.  This defines the
Noether energy as
\begin{equation}
\scripte = \sumsp\int\dL\,f\,H + \int\dV\,{\Bpp^2\over 8\pi}
\label{eqnoethere}
\end{equation}
In the electrostatic case it is simply the integral over $fH$ summed
over species.  
Since the background
magnetic field (through $\vec A$) is not varied, it does not appear in
the Noether energy.

The same follows for the Noether momentum, by applying 
small variations to the space component \cite{LandauLifshitz}.  
In the gyrokinetic case this was given in abstract form by Refs.\
\cite{Sugama00,Brizard00}, but not for specific cases.  Working out the
space components is complicated by the fact that in low frequency
dynamics in a magnetised plasma, the magnetic field serves as an anchor
for momentum, so that the general four-vector version is not conserved
for the dynamics under consideration: not only is the Poynting momentum
neglected against the plasma momentum in
ExB motion (through the assumption of quasineutrality), but also the
neglect of compressional Alfv\'en dynamics removes the exchange with the
background field.  In a tokamak, only toroidal momentum is conserved.
For $L$ of the form given in Eq.\ (\ref{eqaction}),
the conserved toroidal momentum is simply given by the toroidal
canonical momentum weighted by $f$ and summed over species.   
The Noether toroidal momentum is
\begin{equation}
\scriptp = \sumsp\int\dL\,f\,\Pphi
\label{eqnoetherm}
\end{equation}
where $\Pphi=\pt L/\pt\dot\varphi$ and $\varphi$ is the geometric
toroidal angle.  This result is a consequence of all the dependence of
$L$ upon $\varphi$ is in the time component ($H$) or in the field terms
and in the latter there is no time derivative dependence.

One other consideration is that one would like a local form of the
conservation law in terms of a vector momentum density, a symmetric
stress tensor for momentum transport, and a vector describing the
magnetic field anchor explicitly, but this has yet to be worked out.
Herein, we consider energy and toroidal momentum only, and explain their
conservation using the antisymmetric bracket form of the gyrokinetic
equation and the functional derivatives which describe the self
consistent field equations.  One motivation for this is that it is
possible to directly code the antisymmetric bracket form in numerical
simulations, so it is then known that the form of the equations as actually
used is indeed energetically consistent.

Once the Noether energy and toroidal momentum are
known, appropriate operations on the
equations of motion (here, the gyrokinetic equation and the self
consistent field equations) may be used to construct a local form with
the time derivative of an evolving energy/momentum
density and the divergence of an overall transport flux
\cite{Brizard00}.  We will do this herein as well, as part of the
overall motivation to establish the correspondence to fluid and MHD
forms. 

\subsection{Energetic Consistency}

Both of these approaches lead naturally to the known results on
energetic consistency, namely that the same $H$ must be used to obtain
the gyrokinetic equation and the (related) field equations.
Approximations are done in $L$ (hence $H$) and then the equations are
derived without further approximation.  Specifically, orderings in the
derivation of $L$ are used (via Lie transforms or some other method), 
but not thereafter in the derivation of the Euler-Lagrange equations.
Ref.\ \cite{Sugama00} points out specifically that if the drift motion
is to be followed with lowest-order forms of $H$, all higher-order forms
must be cast into field terms.  For example, if $H=H_0+H_1+H_2$, with
the last piece containing the ExB energy, and it is desired to advance
the gyrokinetic equation only with $H_0+H_1$, then the term $fH_2$ in
$L$ must be replaced by $f_0H_2$ where $f_0$ is a background static form
which can be thought of as part of the geometry.  Then, since $H_2$ does
not multiply $f$, it is not involved in the gyrokinetic equation itself 
but only as a field term which would appear on the right hand side of
the polarisation equation, e.g.,
\begin{equation}
\sumsp{\delta \over\delta\phi}f(H_0+H_1) 
= -\sumsp{\delta \over\delta\phi}f_0H_2
\label{eqlinpol}
\end{equation}
This is referred to as {\it linearised polarisation}.  The two
assumptions go together: first order drift motion, and the appearance of
$f_0$ in the polarisation term (the right side of Eq.\ \ref{eqlinpol})
Conversely, if one desires to keep the dependent variable $f$ in this
term, restoring Eq.\ (\ref{eqpolarisation}), then the corresponding
$H_2$ must be kept in the drift motion.  This is the basic statement of
energetic consistency in a total-f global model and the essential
references \cite{Sugama00,Brizard00} arrived at this result ten years
ago.  The same result is found for the same reasons in gyrofluid field
theory models which have a different starting point but are also
Lagrangian/Hamiltonian models \cite{Strintzi04,Strintzi05}.  It is
related to the connection in fluid models between advection and
divergence forms of the equation of motion with respect to the
polarisation drift velocity in a fluid model
(why the polarisation drift must be kept in advection if the species
mass density involves the dependent variable for species particle
density \cite{transport}).

A clear extension of this is that in any discussion of drift motion past
first order, say to order $n$, 
the $f$ must be kept as the dependent variable in all of
the terms $H_0+H_1+\cdots+H_n$ in the functional derivatives in the
polarisation equation.  Any energetic contributions $H_{n+1}+\cdots$
must then either be dropped or combined with a background $f_0$ in $L$,
with the polarisation equation then becoming the appropriately
generalised version of Eq.\ (\ref{eqlinpol}).  Of course, if $L$ is
derived or constructed first and then the Euler-Lagrange equations are
derived without approximation thereafter, then energetic consistency
becomes a guaranteed result.

\subsection{Time Symmetry and Energy Conservation}

Using the antisymmetric bracket form of the gyrokinetic equation
(Eq.\ \ref{eqgk4}) we multiply by $H$ and use the linearity of the
derivatives $Hf_{,b} = (fH)_{,b} - fH_{,b}$ and the antisymmetry (the
form with $H_{,a}H_{,b}$ vanishes due to the permutation of indices in
$\Epsilon^{abcd}$), to find
\begin{equation}
\ptt{}(fH) + \Epsilon^{abcd}H_{,a}(fH)_{,b}A^*_{c,d} = f\ptt{H}
\label{eqlocalenergy}
\end{equation}
which is the local energy equation in phase space.

Integration over phase space and summation over species yields
\begin{equation}
\sumsp\int\dL\,\ptt{}(fH) = \sumsp\int\dL\,f\ptt{H}
\end{equation}
with the bracket vanishing under the integral.
Under the integral the right side is replaced by functional derivatives
(cf.\ Appendix A)
\begin{equation}
\sumsp\int\dL\,\ptt{}(fH) 
= \int\dV\sumsp{\delta fH\over\delta\phi}\ptt{\phi}
+ \int\dV\sumsp{\delta fH\over\delta\Apl}\ptt{\Apl}
\end{equation}
The first term on the right side vanishes, due to polarisation
(Eq.\ \ref{eqpolarisation}).  The second is replaced by the field term
in $\Apl$, due to induction (Eq.\ \ref{eqinduction}), so that
\begin{equation}
\sumsp\int\dL\,\ptt{}(fH) 
= \int\dV {1\over 4\pi}\ddpp\Apl\ptt{\Apl}
\end{equation}
Integration of the divergence operator in $\ddpp$ by parts then yields
\begin{equation}
\sumsp\int\dL\,\ptt{}(fH) 
= - \int\dV {1\over 4\pi}\dpp\Apl\cdot\ptt{}\dpp\Apl
= - \int\dV {1\over 8\pi}\ptt{}\abs{\dpp\Apl}^2
\end{equation}
Identification with $\Bpp^2$ in Eq.\ (\ref{eqbpp}) then recovers
\begin{equation}
\ptt{}\LP\sumsp\int\dL\,f\,H + \int\dV\,{\Bpp^2\over 8\pi}\RP = 0
\label{eqecons}
\end{equation}
which is the same as obtained using the Noether theorem
(cf.\ Eq.\ \ref{eqnoethere}).
This is the statement of energy conservation and it is valid
under time symmetry for any dependence of $H$ upon $\phi$ and $\Apl$
given the form of $L$ stated above.

The dependence of energy conservation upon time symmetry is contained in
the step from the time derivative to the functional derivative, as
$\ppt{}$ must commute with any of the differential operators
involved in the dependence of $H$ upon $\phi$ and $\Apl$.  The
requirement of energetic consistency is evident in the 
fact that the same $H$ is used
in the gyrokinetic equation as that whose functional derivatives appear
in the polarisation and induction equations.  Also, the same $f$ must
appear with each of the terms in $H$ in all cases, or else the symmetry
is broken.

\subsection{Axisymmetry and Toroidal Momentum Conservation}

In this case the properties are different.  For energy, $H$ depends on
both time and toroidal angle, but the symmetry of the bracket allowed
combination of $(fH)$ there.  Canonical momentum at the particle level
is given by
\begin{equation}
\Pphi = {e\over c}\Aphi = {e\over c}A_\varphi + p_z\bphi
\label{eqpphi}
\end{equation}
This form is both static and axisymmetric, but does not appear in the
bracket.  We multiply by $\Pphi$ to find
\begin{equation}
\ptt{}(f\Pphi) + \Epsilon^{abcd}H_{,a}(f\Pphi)_{,b}A^*_{c,d} 
= f{e\over c}\Epsilon^{abcd}H_{,a}A^*_{\varphi,b} A^*_{c,d} 
\end{equation}
Because the two appearances of $A^*$ in the right side appear with
different indices, we may make some symmetry arguments.  First, the
indices $b$ and $d$ cannot be $\varphi$, due to axisymmetry.  Second,
index $c$ cannot be $\varphi$, or else $\Aphi$ appears twice and
the remaining permutation over
indices $\{abd\}$ vanishes.  Hence, index $a$ must be $\varphi$ and the
others are among the other three coordinates, so that
\begin{equation}
\ptt{}(f\Pphi) + \Epsilon^{abcd}H_{,a}(f\Pphi)_{,b}A^*_{c,d} 
= f\pvpvp{H}{e\over c}\Epsilon^{\varphi abc}A^*_{\varphi,a} A^*_{b,c} 
\label{eqfpphi}
\end{equation}
Of the remaining terms, index $b$ cannot be $z$ because there is no
$A^*_z$, so in each term one of $a$ or $c$ is $z$ while the others are
the two coordinates covering the perpendicular plane, which we can label
$1$ and $2$ (hence $dx^3$ in $\dV$ is $d\varphi$).  We also observe that
\begin{equation}
\ppzpz{\Astar} = {c\over e}\bunit
\end{equation}
which cancels the $(e/c)$ factor.  Noting that the units of
$\Epsilon^{abcd}$ are $1/\sqrt{g}\Bpl$, we set the positive permutation
as $\{\varphi\,z\,1\,2\}$ and permute the $\{z\,1\,2\}$ indices to find
\begin{equation}
{e\over c}\Epsilon^{\varphi abc}A^*_{\varphi,a} A^*_{b,c} 
= {1\over\sqrt{g}\Bpl}\LB b_{\varphi}(A^*_{1,2} - A^*_{2,1})
+ b_{1}(A^*_{2,\varphi} - A^*_{\varphi,2})
+ b_{2}(A^*_{\varphi,1} - A^*_{1,\varphi})\RB
\end{equation}
where we have eliminated the two (zero) terms 
$A^*_{\varphi,1}A^*_{z,2}$ and
$A^*_{\varphi,2}A^*_{z,1}$, and replaced them with the two (zero) terms
$b_{2}A^*_{1,\varphi}$ and $b_{1}A^*_{2,\varphi}$, respectively.
We observe that
\begin{equation}
{1\over\sqrt{g}}\LB b_{\varphi}(A^*_{2,1} - A^*_{1,2})
+ b_{1}(A^*_{\varphi,2} - A^*_{2,\varphi})
+ b_{2}(A^*_{1,\varphi} - A^*_{\varphi,1})\RB \equiv
\bunit\cdot\curl\Astar = \Bpl
\end{equation}
and note the switch in the order of coefficients to the expression just
above.  Therefore we have
\begin{equation}
{e\over c}\Epsilon^{\varphi abc}A^*_{\varphi,a} A^*_{b,c} = -1
\end{equation}
Putting this into the right side of Eq.\ (\ref{eqfpphi}), we find
\begin{equation}
\ptt{}(f\Pphi) + \Epsilon^{abcd}H_{,a}(f\Pphi)_{,b}A^*_{c,d} = -f\pvpvp{H}
\label{eqlocalmomentum}
\end{equation}
which is the local toroidal momentum equation in phase space.

This is the same form as the result obtained from a canonical
representation of the particle Lagrangian
\begin{equation}
L_p = P_\psi\dot\psi + P_\theta\dot\theta + \Pphi\dot\varphi 
+ {mc\over e}\mu\thetadot - H
\end{equation}
written directly in terms of the coordinates $\{\psi\,\theta\,\varphi\}$
as was once usual \cite{WhiteChance84}).
The corresponding 
Euler-Lagrange equation 
for the toroidal angle 
is
\begin{equation}
\dot\Pphi = -\pvpvp{H}
\end{equation}
due to the axisymmetry of the rest of $L_p$.
Using the advection forms
\begin{equation}
\dot f = \ptt{f} + \zpdot\cdot\pzzp{f} = 0
\qqquad
\dot \Pphi = \ptt{\Pphi} + \zpdot\cdot\pzzp{\Pphi} = -\pvpvp{H}
\end{equation}
we find
\begin{equation}
\ptt{}(f\Pphi) + \zpdot\cdot\pzzp{}(f\Pphi) = -f\pvpvp{H}
\end{equation}
which is the same form as in Eq.\ (\ref{eqlocalmomentum}) with the
bracket recast in terms of an advection.

Returning to Eq.\ (\ref{eqlocalmomentum}),
integration over phase space and summation over species yields
\begin{equation}
\sumsp\int\dL\,\ptt{}(f\Pphi) = -\sumsp\int\dL\,f\pvpvp{H}
\end{equation}
with the bracket vanishing under the integral.
Under the integral the right side is replaced by functional derivatives
(cf.\ Appendix A)
\begin{equation}
\sumsp\int\dL\,\ptt{}(f\Pphi) 
= \int\dV\sumsp{\delta fH\over\delta\phi}\pvpvp{\phi}
+ \int\dV\sumsp{\delta fH\over\delta\Apl}\pvpvp{\Apl}
\end{equation}
The manipulations follow the energy derivation, with $\ppt{}$ replaced
by $\pt/\pt\varphi$.  In this case the derivative of $\Bpp^2$ with
respect to $\varphi$ vanishes under the integral, so that
\begin{equation}
\ptt{}\sumsp\int\dL\,f\,\Pphi = 0
\label{eqmcons}
\end{equation}
which is the same as obtained using the Noether theorem
(cf.\ Eq.\ \ref{eqnoetherm}).
This is the statement of toroidal momentum conservation and it is valid
under axisymmetry for any dependence of $H$ upon $\phi$ and $\Apl$
given the form of $L$ stated above.

The dependence of toroidal momentum conservation upon axisymmetry is
contained in 
the step from the toroidal angle derivative to the functional derivative, as
$\pt/\pt\varphi$ must commute with any of the differential operators
involved in the dependence of $H$ upon $\phi$ and $\Apl$.  The
requirement of energetic consistency is evident in the same way as for
energy with the same loss of consistency if the symmetry between
functional derivatives and the gyrokinetic equation is broken.

\subsection{Phase Space
Continuity Forms of Energy and Toroidal Momentum Conservation}

Eqs.\ (\ref{eqlocalenergy},\ref{eqlocalmomentum}) give the antisymmetric
bracket forms of the local energy and toroidal momentum equations in
phase space.  Using toroidal momentum as an example,
We identify
\begin{equation}
\Epsilon^{abcd}H_{,a}f_{,b}A^*_{c,d} = \zpdot\cdot\pzzp{f}
\end{equation}
Then, the phase space conservation condition
(Eq.\ \ref{eqphasespace}) can be used to express
Eq.\ (\ref{eqlocalmomentum}) as
\begin{equation}
\ptt{}(f\Pphi) + {1\over\sqrt{g}\Bpl}
\pzzp{}\cdot(\sqrt{g}\Bpl\,f\Pphi\,\zpdot) = -f\pvpvp{H}
\label{eqmomentumcontinuity}
\end{equation}
which is the phase space continuity equation for toroidal momentum.

Similarly, we may express Eq.\ (\ref{eqlocalenergy}) as
\begin{equation}
\ptt{}(fH) + {1\over\sqrt{g}\Bpl}
\pzzp{}\cdot(\sqrt{g}\Bpl\,fH\,\zpdot) = f\ptt{H}
\label{eqenergycontinuity}
\end{equation}
which is the phase space continuity equation for energy.

Eq.\ (\ref{eqmomentumcontinuity}) makes it obvious that $f\Pphi$
is conserved locally in axisymmetric systems ($\pt/\pt\varphi=0$), and
globally in any geometry.  In Eq.\ (\ref{eqenergycontinuity}) the term
appears instead with $\ppt{H}$, which eventually accounts for the
magnetic energy as in Eq.\ (\ref{eqecons}).
We will use these continuity
equations to produce local transport equations for
both momentum and energy later.

\section{A Comment on Ordering}

Before the emergence of the Lagrangian/Hamiltonian approach to
drift kinetic particle motion 
\cite{Littlejohn81,Littlejohn83,WhiteChance84}, in which the equation
for $f$ is built explicitly using the
phase space positions of the gyrocenters as dependent variables,
it was customary to start with the Vlasov (or Boltzmann) equation for
the particles themselves and apply gyroaveraging through a successive
ordering \cite{Catto81,FriemanChen82}.  This causes problems, however,
if applied na\"\i vely (and strictly) to the field equations.
Considering an electrostatic model ($\Apl=0$), we may expand $H$ in
terms of small amplitude fluctuations ($e\phi/T_e\sim\rho_* =
\rho_i/\Lpp$, where $\rho_i$ is the thermal ion gyroradius and $\Lpp$ is
the profile gradient scale length \cite{Taylor68,Rutherford68})
\begin{equation}
H = H_0 + H_1 + H_2 + \cdots + H_n
\end{equation}
where at each order $n$ the term $H_n$ is $n$-th order in $\phi$.
As long as $f$ is not expanded order by order, there is no problem.  The
same $f$ multiplies each $H_n$ in turn and due to the linearity property
the functional derivatives add, producing Eq.\ (\ref{eqpolarisation})
term by term.

However, if $f$ is also ordered such that $f_0$ is the background
(usually a Maxwellian) and $f_1$ is the fluctuation, then there is a
problem.  Recall that if drift motion is included to order $n$ in $H$
through $\phi$ then the polarisation equation must include $fH_n$ in the
functional derivatives to preserve energetic consistency.  The $f$ must
include both $f_0$ and $f_1$.  However, if the ordering is truncated at
order $n$ then the term 
$\delta(f_1H_n)/\delta\phi$ is missing.  Formally, it is order
$n+1$.  So this $(n+1)$-th order term must be kept, but in doing so we
violate the ordering scheme.  If the ordering scheme is applied to
expand $f$ and keep all terms up to order $n$, dropping all order $n+1$
terms, then this one piece will be
missing.  This problem is present at any order of expansion, at the last
order.  The only acceptable solution for orderings
is to expand $H$ in orders but not $f$.  That
is, polarisation is not to be linearised (cf.\ Eq.\ \ref{eqlinpol}) if
contributions above linear order in $H$ to the drift motion are
considered.  Hence any discussion of orderings in which higher order
drifts (even second order) are considered should be done under full
energetic consistency.  As noted, the field theory version of
gyrokinetics is the only straightforward way to guarantee this.

\section{The ExB Vorticity Transport Equation
\label{sec:vorticity}}

We may form an equation for the gyrocenter charge density by multiplying
the gyrokinetic equation by the charge $e$ for each species
and summing over species.
For general dependence of $H$ upon $\phi$, 
the terms linear in $\phi$ are collected and all the
others are combined into a total divergence.  We may separate
\begin{equation}
H = H_0 + e\phi + H_P
\label{eqformh}
\end{equation}
where $H_0$ comprises all the terms not involving $\phi$ and then the
polarisation piece $H_P$ may be constructed from $H-H_0-e\phi$.  All
gyroaveraging corrections (e.g., from $1-J_0$) are collected into
$H_P$.  We define the generalised vorticity $\Omega$ and the
polarisation vector $\vec P$ such that
\begin{equation}
\div\vec P \equiv -\Omega \equiv \sumsp fe 
\label{eqvorticity}
\end{equation}
The quantity on the right side is the gyrocenter charge density.  Since
the derivations are being done under strict quasineutrality
($E^2/8\pi$) was neglected as discussed around Eq.\ \ref{eqaction}), the
quantity on the left side balances this.  It is the polarisation density
as developed by other methods in Ref.\ \cite{Lee83}.  The generalised
vorticity is defined in this manner as a quantity sensitive to small
scales which (as seen below) in the MHD limit reduces to the simple ExB
vorticity. 

Then the polarisation equation (Eq.\ \ref{eqpolarisation}) is
\begin{equation}
\div\vec P = - \sumsp{\delta fH_P\over\delta\phi} 
\label{eqpolvector}
\end{equation}
It is essential for the subsequent results to be able
to write the polarisation equation in this form, with polarisation
density given by a divergence.

For any $H$ with dependence on $\phi$ such that the separation
$H = H_0 + e\phi + H_P$ yields dependence of $H_P$ upon $\phi$ only
through $\grad\phi$ and $\ddpp\phi$, the functional derivative of $fH_P$
is
\begin{equation}
{\delta fH_P\over\delta\phi} = 
\int\dW\LB\ddpp \LP f{\pt H_P\over\pt\ddpp\phi}\RP
- \div \LP f{\pt H_P\over\pt\grad\phi}\RP\RB
\end{equation}
which can be written as a divergence
\begin{equation}
{\delta fH_P\over\delta\phi} = 
\div\int\dW\LB 
\dpp\LP f{\pt H_P\over\pt\ddpp\phi}\RP - f{\pt H_P\over\pt\grad\phi} 
\RB
\end{equation}
and then the species sum of the
quantity in the square brackets is identified with 
$\vec P$.  Since within $H$ only $H_P$ depends on $\grad\phi$ or
$\ddpp\phi$ we generally have
\begin{equation}
\vec P = \sumsp\int\dW\LB
f{\pt H\over\pt\grad\phi} - \dpp\LP f{\pt H\over\pt\ddpp\phi}\RP\RB
\label{eqpolform}
\end{equation}
and the need to be able to write
Eq.\ (\ref{eqpolvector}) in that form is satisfied.  The only
restriction on the form of $H$ is that all terms past first order in
$\phi$ appear only through $\grad\phi$ or $\ddpp\phi$.  Note, however,
that there is no such restriction on $\Apl$.

For example, for the
long-wavelength electrostatic $H$ through second order,
\begin{equation}
H = {p_z^2\over 2m} + \mu B + e\phi - {mc^2\over 2B^2}\abs{\dpp\phi}^2
\end{equation}
the polarisation equation is
\begin{equation}
\sumsp\int\dW\LB e f + {1\over\Bpl}\div\Bpl{fmc^2\over B^2}\dpp\phi\RB = 0
\end{equation}
and the polarisation vector is
\begin{equation}
\vec P = - \sumsp\int\dW\, {fmc^2\over B^2}\dpp\phi
\end{equation}
where we note that $\dW/\Bpl$ commutes with spatial derivatives.  
In this case the species-summed 
velocity space integral is straightforward and we have
\begin{equation}
\vec P = - \rho_M{c^2\over B^2}\dpp\phi
\qqquad
\Omega = \div\rho_M{c^2\over B^2}\dpp\phi
\end{equation}
where $\rho_M$ is the species sum of $nm$ with $n=\int\dW f$ the species
density.
One sees why the gyrocenter charge density plays the role of a
(negative) vorticity, since $\div\vec P$ is proportional to $-\ddpp\phi$
plus corrections due to the gradients of the densities and the magnetic
field strength.

The global conservation is trivial since the phase space integral of the
gyrokinetic equation conserves particles for each species.  The
vorticity transport equation may be written as
\begin{equation}
\ptt{}\avg{\Omega} - \pVV{}\avg{fe\dot V} = 0
\label{eqvortrans}
\end{equation}
Here, the angle brackets denote the flux surface average, which is the
same as the volume derivative of an integral over the volume enclosed by
a particular flux surface \cite{HintonHazeltine76}.  
For simplicity we assume Hamada flux coordinates $\{V\theta\zeta\}$
where $V$ is the volume enclosed by the particular flux surface,
the contravariant components of the magnetic field are functions of $V$
only, and the poloidal and toroidal angles (respectively)
are unit-cycle periodic \cite{Hamada58,Hamada62,HintonHazeltine76}.
This leaves $\sqrt{g}=1$.  Then,
$\dot V=\Rdot\cdot\grad V$ is the contravariant $V$-component
of $\Rdot$.

We have used the
property that the flux surface average of a phase space divergence
annihilates the velocity coordinate derivatives (with respect to $p_z$
and $\mu$) and commutes the integration $\int\dW/\Bpl$ past the spatial
derivatives.  Then, the flux surface average annihilates the angle
derivatives in a flux coordinate representation, leaving the
$V$-component of the drift motion and the derivative $\pt/\pt V$.  Note
that the flux surface average of a kinetic quantity implies the
species-summed velocity space integration, which is left understood.

The introduction of
the polarisation vector under the time derivative leaves this equation
as a pure divergence,
\begin{equation}
\pVV{}\avg{\ptt{P^V} + fe\dot V} = 0
\label{eqvorcancel}
\end{equation}
where the superscript $V$ denotes the contravariant $V$-component.  This
describes charge conservation in the form $\div\vec J=0$, as a balance
between the gyrocenter drift current (the species sum of $fe\Rdot$,
including the parallel piece) and
the divergence of $\vec P$, so we may also identify
$\ppt{\vec P}$ as the polarisation current.
The quantity given by the flux surface average in
Eq.\ (\ref{eqvorcancel}), which includes both pieces,
can be taken to vanish
everywhere given appropriate boundary conditions (e.g., it vanishes at
the magnetic axis, where $V=0$, due to regularity of the vector
component).  Hence we will also have
\begin{equation}
\pVV{}\gamma(V)\avg{\ptt{P^V} + fe\dot V} = 0
\label{eqvorcancelb}
\end{equation}
where $\gamma$ is any flux surface quantity (also called flux function).
Specifically, this expression vanishes for $\gamma=A_\varphi$ since
$A_\varphi$ is the quantity whose isosurfaces define flux surfaces, for
a tokamak magnetic field.

\section{A Further Comment on Ordering
\label{sec:fluxtube}}

In the conventional gyrokinetic ordering the small parameter is equivalently
$\kpl/\kpp$ in the wavenumber anisotropy or the fluctuation
amplitude $e\phi/T_e$,
which are used interchangeably \cite{Taylor68,Rutherford68}.  
(The often stated ordering of $\rho_*\ll 1$ is actually {\it a
  posteriori} as it follows from the ultimate requirement that the
resulting dynamics is in the range of the
diamagnetic frequency $\omega_*$, and for this to be small compared to 
the ion gyrofrequency requires $\rho_*\ll 1$.)
Under
these conditions we may simplify expressions by using the field aligned
version \cite{Dewar83,fluxtube}\
of Hamada flux coordinates \cite{Hamada58,Hamada62,HintonHazeltine76}.
This discussion
follows the version used in Ref.\ \cite{fluxtube} which includes the
definitions and construction algorithms for the coordinates.
Starting with $\{V\theta\zeta\}$ as above,
we transform the
toroidal angle only, defining $\xi=\zeta-q\theta$ where $q=q(V)$ is a
flux function giving the ratio $B^\zeta/B^\theta$ in the contravariant
components.  Then, both $B^V$ and $B^\xi$ vanish, and the only
nonvanishing component of $B$ is $B^\theta$.  This is defined as 
$B^\theta=\chiv\equiv\ppV{\chi}$, where $\chi=\chi(V)$ is another flux
definition.  The tokamak magnetic field
\begin{equation}
\vec B = I\grad\varphi + \grad A_\varphi\cross\grad\varphi
\label{eqdefb}
\end{equation}
may be written as
\begin{equation}
\vec B = \grad\xi\cross\grad\chi = \chiv\grad\xi\cross\grad V
\end{equation}
which is called a Clebsch representation \cite{Dewar83}.  The sign
conventions are $\grad R\cross\grad Z\cdot\grad\varphi>0$ and
$\chiv>0$.  It follows that $\chiv=-2\pi\,\ppV{\psi}$ and that for any
vector (including the gradient)
the covariant components satisfy $A_\xi=A_\zeta=A_\varphi/2\pi$.  With
only $B^\theta$ nonvanishing, $\pph{}$ tracks the parallel derivative
and therefore we may assume $\pph{}\ll\ppV{},\ppxi{}$ due to the
wavenumber ordering.

We consider the flux surface average of the
toroidal momentum continuity equation
(Eq.\ \ref{eqmomentumcontinuity})
\begin{equation}
\cdots + \pVV{}\avg{f{e\over c}A_\varphi 
\LP\grad H_f\cdot{c\over e}{\vec F\over B\Bpl}\RP^V} = -\avg{f\pvpvp{H_f}}
\label{eqdrifts}
\end{equation}
where the focus is on the term in the drift motion involving
$A_\varphi/c$, and $H_f$ is the part of $H$ which involves the
fluctuations in the field variables.  We may apply the flute mode
ordering to the drifts term and assume $\Bpl=B+O(\rho_*)$ dropping the
small correction, so that
\begin{equation}
\LP\grad H_f\cdot{c\over e}{\vec F\over B\Bpl}\RP^V 
= {c\over e}{1\over\chiv}\pxixi{H_f}
\end{equation}
where we have used the exact equality $F^{\xi V}/B^2=1/\chiv$ which
results from the index-raising operations on $F_{\xi V}=\chiv$ and the
equality $(B/\chiv)^2=g^{\xi\xi}g^{VV}-g^{\xi V}g^{V\xi}$ giving
the determinant of the perpendicular contravariant metric coefficients.
Recall that $\sqrt{g}=1$ and $F_{\xi V}=\epsilon_{\xi V\theta}B^\theta$.
Substituting the drifts into Eq.\ (\ref{eqdrifts}),
the factors of $(c/e)$ cancel and we find
\begin{equation}
\cdots + \pVV{}{A_\varphi\over\chiv}\avg{f \pxixi{H_f}} = -\avg{f\pvpvp{H_f}}
\end{equation}
where the flux quantities have been taken out of the flux surface
average.  Finally we insert the $2\pi$ normalisation in $\xi$ so that
\begin{equation}
\cdots + \pVV{} {2\pi A_\varphi\over\chiv}\avg{f \pvpvp{H_f}}
= -\avg{f\pvpvp{H_f}}
\end{equation}
We observe that the two flux surface averages are the same term.
Furthermore, for a local model
we may evaluate $A_\varphi=A_{\varphi,V}(V-V_0)$ near a
zero at $V=V_0$ because the 
amplitude of the flux is arbitrary, and use $2\pi A_{\varphi,V}=-\chiv$,
so that
\begin{equation}
\cdots + \pVV{}\LB (V_0-V)\avg{f \pvpvp{H_f}}\RB = -\avg{f\pvpvp{H_f}}
\label{eqordering}
\end{equation}
where it is evident that the two terms are at the same order for any
contribution by $\phi$ or $\Apl$ to $H$.

It is tempting to draw the cancellation obtained by differentiating the
$V_0-V$ factor.  However, the remnant is of the same order.  Though this
manipulation does not yield a useful final result it still serves to
show that the potentially large term involving $A_\varphi/c$ gives a
contribution at the same order as the terms we keep on the right side of 
Eq.\ (\ref{eqmomentumcontinuity}).  Again, applying ordering to $H$
only, the similarity in magnitude follows order by order, as the $H_n$
contributions on each side are similar at each order.  This dispenses
with any notion that the large factor $A_\varphi/c$ should be
accompanied by higher order drift terms due to $H_f$ than are present in
the term $-\avg{fH_{,\varphi}}$ on the right side.  We will find it
necessary to keep the second order ExB energy term to obtain useful
results, but since energetic consistency and the consistency in
ordering we have just derived indicate, we do not need to keep terms 
beyond second order in $\phi$ and $\Apl$ on the left side of 
Eq.\ (\ref{eqmomentumcontinuity}).  Again, the underlying energetic
consistency yields straightforward conclusions consistent with the use
of a single $H$ (to any order) everywhere within any particular version
of the gyrokinetic field theory model.

The dependence of this result on energetic consistency cannot be
overemphasised.  In an ordering expansion, $A_\varphi/c$ as a large term
potentially introduces higher order terms in $H$.  However, application
of the field equations (through functional derivatives in $H$) up to any
particular order, which obey exact energetic consistency up to the
same order, was used to arrive at the result that the two expressions
involving $\avg{fH_{,\varphi}}$ are at the same order (each contribution to
$H$, order by order).  Had the field equations been missing terms at the
highest order kept in the drifts in the right side of
Eq.\ (\ref{eqmomentumcontinuity}), this result would not have been
obtained and we would be required to discuss spurious effects.  This is
why discussion of orderings in $H$ past first order must be done in the
context of energetic consistency.

\section{The Toroidal Momentum Transport Equation
\label{sec:momentum}}

We now do for toroidal momentum what we did for vorticity in
Sec.\ \ref{sec:vorticity}. 
Starting with Eq.\ (\ref{eqmomentumcontinuity}), we apply species
summation and velocity space integration and then the flux surface
average to obtain
\begin{equation}
\ptt{}\avg{f\Pphi} + \pVV{}\avg{f\Pphi\dot V} = -\avg{f\pvpvp{H}}
\end{equation}
where as before $\dot V=\Rdot\cdot\grad V$, the velocity space
integration annihilates velocity space derivatives and $\dW/\Bpl$
commutes past the spatial divergence.  We insert the definition of
$\Pphi$ in Eq.\ (\ref{eqpphi}) and collect the $A_\varphi$ terms to find
\begin{equation}
\ptt{}\avg{f{e\over c}A_\varphi} + \pVV{}\avg{f{e\over c}A_\varphi\dot V} 
+ \ptt{}\avg{fp_zb_\varphi} + \pVV{}\avg{fp_zb_\varphi\dot V} = -\avg{f\pvpvp{H}}
\end{equation}
We pull $A_\varphi$ out of the flux surface average to find
\begin{equation}
{A_\varphi\over c}\ptt{}\avg{fe} + \pVV{}{A_\varphi\over c}\avg{fe\dot V} 
+ \ptt{}\avg{fp_zb_\varphi} + \pVV{}\avg{fp_zb_\varphi\dot V} = -\avg{f\pvpvp{H}}
\label{eqmomentuma}
\end{equation}
Using Eqs.\ (\ref{eqvorticity},\ref{eqpolvector}),
the gyrocenter charge density is replaced by the vorticity under the
time derivative, to find
\begin{equation}
{A_\varphi\over c}\ptt{}\avg{\div\vec P} + \pVV{}{A_\varphi\over c}\avg{fe\dot V} 
+ \ptt{}\avg{fp_zb_\varphi} + \pVV{}\avg{fp_zb_\varphi\dot V} = -\avg{f\pvpvp{H}}
\end{equation}
The divergence operator on $\vec P$ is done by parts to find
\begin{eqnarray}
& & - \ptt{}\avg{{1\over c}\vec P\cdot\grad A_\varphi}
+ \pVV{}{A_\varphi\over c}\ptt{}\avg{P^V} 
+ \pVV{}{A_\varphi\over c}\avg{fe\dot V} 
\nonumber\\ & & \qquad {} 
+ \ptt{}\avg{fp_zb_\varphi} + \pVV{}\avg{fp_zb_\varphi\dot V} 
= -\avg{f\pvpvp{H}}
\end{eqnarray}
since $\grad=\grad V(\ppV{})$ for any flux function.  
In pulling the divergence out of the flux surface
average we have used $\avg{\div\vec P}=\ppV{}\avg{P^V}$.  We combine the
first two divergence terms to find
\begin{equation}
\pVV{}{A_\varphi\over c}\ptt{}\avg{P^V} 
+ \pVV{}{A_\varphi\over c}\avg{fe\dot V} 
= \pVV{}{A_\varphi\over c}\avg{\ptt{P^V} + fe\dot V} = 0
\end{equation}
which vanishes by Eqs.\ (\ref{eqvorcancel},\ref{eqvorcancelb}).  
Hence we have
\begin{equation}
{A_\varphi\over c}\ptt{}\avg{fe} + \pVV{}{A_\varphi\over c}\avg{fe\dot V} 
= - \ptt{}\avg{{1\over c}\vec P\cdot\grad A_\varphi}
\label{eqaplcancel}
\end{equation}
as a result for any $H$ subject to the comment after Eq.\ (\ref{eqpolform}), 
through the dependence of the vorticity equation (Eq.\ \ref{eqvorcancel})
upon the polarisation equation (Eq.\ \ref{eqpolarisation}).
Insertion of Eq.\ (\ref{eqaplcancel})
into Eq.\ (\ref{eqmomentuma}) and moving the right side term to the left
side produces
\begin{equation}
- \ptt{}\avg{{1\over c}\vec P\cdot\grad A_\varphi}
+ \ptt{}\avg{fp_zb_\varphi} + \pVV{}\avg{fp_zb_\varphi\dot V} 
+ \avg{f\pvpvp{H}} = 0
\label{eqmomentumb}
\end{equation}
This result eliminates terms in which $A_\varphi$, the flux label,
appears by itself (not under a gradient operator).
We may identify the first and
second terms as the time derivative of a total toroidal momentum density
consisting of an ExB part and a parallel part (both have toroidal
components).  The next term gives the drift effects of the parallel part
of the toroidal momentum density.  The last term has already been seen
to vanish under the total phase space integral, so as a flux surface
average it should be another (set of) transport divergence term(s).  It
remains to show this.

\subsection{Field Terms for Simple Hamiltonian Dependences}

We assume at first for illustration purposes an electrostatic case with
general dependence of $H$ upon $\phi$ and $\grad\phi$ irrespective of
any ordering. 
We have the integrand in the last term of Eq.\ (\ref{eqmomentuma}) as 
\begin{equation}
f\pvpvp{H}=f{\pt H\over\pt\phi}\pvpvp{\phi}
+ f{\pt H\over\pt\grad\phi}\cdot\grad\pvpvp{\phi}
\end{equation}
noting that the partial derivative and the gradient
commute.  The gradient is done by parts to find
\begin{equation}
f\pvpvp{H}
= \LP f{\pt H\over\pt\phi}-\div f{\pt H\over\pt\grad\phi}\RP\pvpvp{\phi}
+ \div \LP \pvpvp{\phi}\,f{\pt H\over\pt\grad\phi}\RP
\end{equation}
The first expression in parentheses is the functional derivative and it
vanishes under the eventual species-summed velocity space integration,
due to Eq.\ (\ref{eqpolarisation}).  The second term gives a
divergence of a transport flux.  Putting this back under the flux
surface average, we have
\begin{equation}
\avg{f\pvpvp{H}} = \pVV{}
\avg{\pvpvp{\phi}\grad V\cdot f{\pt H\over\pt\grad\phi}}
\end{equation}
which is the desired result.

Following the same analysis for a general dependence of $H$ upon $\phi$
and $\Apl$, with the functional derivative with respect to $\Apl$
not vanishing but replaced by
the field term as per Eq.\ (\ref{eqinduction}) we find
\begin{equation}
\avg{f\pvpvp{H}}=
\pVV{}\avg{\pvpvp{\phi}\grad V\cdot f{\pt H\over\pt\grad\phi}}
+ \pVV{}\avg{\pvpvp{\Apl}\grad V\cdot 
	\LP f{\pt H\over\pt\grad\Apl} + {1\over 4\pi}\dpp\Apl\RP}
\label{eqfgradh}
\end{equation}
where the contribution due to $\pt\Bpp^2/\pt\varphi$ vanishes under the
flux surface average.
As we will see when explicitly showing the version of the result using
the long-wavelength ``MHD Hamiltonian'' these two terms generally 
give the Reynolds and Maxwell stresses, respectively.

\subsection{Field Terms for Hamiltonian Dependences involving the
  Laplacian
\label{sec:laplacians}}

In the case that $H$ depends on the field variables through not only
$\phi$ and $\grad\phi$ but also $\ddpp\phi$ then there is more to do but
generalisation is straightforward (for background see the text by
Gelfand and Fomin \cite{GelfandFomin}).  We expand
\begin{equation}
f\pvpvp{H}=f{\pt H\over\pt\phi}\pvpvp{\phi}
+ f{\pt H\over\pt\grad\phi}\cdot\grad\pvpvp{\phi}
+ f{\pt H\over\pt\ddpp\phi}\ddpp\pvpvp{\phi}
\end{equation}
again assuming the differential operator commutators vanish.  The first
two terms are done as before but the perpendicular Laplacian involves
two integrations by parts.  The Laplacian piece becomes
\begin{equation}
f{\pt H\over\pt\ddpp\phi}\ddpp\pvpvp{\phi} =
\div f{\pt H\over\pt\ddpp\phi}\dpp\pvpvp{\phi} 
	- \dpp\pvpvp{\phi}\cdot\dpp f{\pt H\over\pt\ddpp\phi}
\end{equation}
Both of these pieces further expand according to
\begin{equation}
\div f{\pt H\over\pt\ddpp\phi}\dpp\pvpvp{\phi} 
	= \div \dpp\LP \pvpvp{\phi}\, f{\pt H\over\pt\ddpp\phi}\RP
	- \div \LP \pvpvp{\phi}\, \dpp f{\pt H\over\pt\ddpp\phi}\RP
\end{equation}
and
\begin{equation}
- \dpp\pvpvp{\phi}\cdot\dpp f{\pt H\over\pt\ddpp\phi}
	= - \div\LP\pvpvp{\phi}\dpp f{\pt H\over\pt\ddpp\phi}\RP
	+ \pvpvp{\phi}\ddpp f{\pt H\over\pt\ddpp\phi}
\end{equation}
The Laplacian piece is then
\begin{eqnarray}
& & f{\pt H\over\pt\ddpp\phi}\ddpp\pvpvp{\phi}
= \pvpvp{\phi}\ddpp f{\pt H\over\pt\ddpp\phi}
\nonumber\\ & & \qquad {}
+\div\LB\dpp\LP \pvpvp{\phi}\, f{\pt H\over\pt\ddpp\phi}\RP
	- 2\LP \pvpvp{\phi}\, \dpp f{\pt H\over\pt\ddpp\phi}\RP\RB
\end{eqnarray}
Putting in the terms from the $\phi$ and $\grad\phi$ dependences, we
have
\begin{eqnarray}
 & & f\pvpvp{H}=\pvpvp{\phi}\LP f{\pt H\over\pt\phi}
	- \div f{\pt H\over\pt\grad\phi}
	+ \ddpp f{\pt H\over\pt\ddpp\phi}\RP
\nonumber\\ & & \qquad {} 
+ \div\LB\pvpvp{\phi}\,\LP f{\pt H\over\pt\grad\phi}
	- 2\dpp f{\pt H\over\pt\ddpp\phi}\RP
+ \dpp\LP\pxixi{\phi}\,f{\pt H\over\pt\ddpp\phi}\RP\RB
\end{eqnarray}
The first term in parentheses gives the functional derivative (which
vanishes according to Eq.\ \ref{eqpolarisation}) and the
rest become transport fluxes.  The flux surface average with
species-summed velocity space integration then yields
\begin{equation}
\avg{f\pvpvp{H}} = \pVV{}\avg{ \grad V\cdot\LB
\pvpvp{\phi}\,\LP f{\pt H\over\pt\grad\phi} 
	- 2\dpp f{\pt H\over\pt\ddpp\phi}\RP
+ \dpp\LP \pvpvp{\phi}\,f{\pt H\over\pt\ddpp\phi}\RP\RB}
\end{equation}
which is the desired result.

Following the same analysis for a general dependence of $H$ upon $\phi$
and $\Apl$, with the functional derivative with respect to $\Apl$
not vanishing but replaced by
the field term as per Eq.\ (\ref{eqinduction}) we find
\begin{eqnarray}
 & & \avg{f\pvpvp{H}} = \pVV{}\avg{ \grad V\cdot\LB
\pvpvp{\phi}\,\LP f{\pt H\over\pt\grad\phi} 
	- 2\dpp f{\pt H\over\pt\ddpp\phi}\RP
+ \dpp\LP \pvpvp{\phi}\,f{\pt H\over\pt\ddpp\phi}\RP\RB}
\nonumber\\ & & \qquad {}
+ \pVV{}\left\langle \grad V\cdot\LB
\pvpvp{\Apl}\,\LP {1\over 4\pi}\dpp\Apl + f{\pt H\over\pt\grad\Apl}
	- 2\dpp f{\pt H\over\pt\ddpp\Apl}\RP
\right.
\right.
\nonumber\\ & & \qqquad {} + \left.\left.
\dpp\LP \pvpvp{\Apl}\,f{\pt H\over\pt\ddpp\Apl}\RP\RB\right\rangle
\label{eqfgradha}
\end{eqnarray}
where the contribution due to $\pt\Bpp^2/\pt\varphi$ vanishes under the
flux surface average.

Though more complicated, this result shows that for general dependence
of $H$ upon $\phi$ and $\Apl$ and their gradients and Laplacians,
all terms in the
toroidal momentum transport equation may be recast as divergences of
flux surface-averaged transport fluxes, and no terms with $A_\varphi$
not under gradients appear.

\subsection{Illustration using an MHD Hamiltonian
\label{sec:mhd}}

We may dispense with finite gyroradius effects while studying
equilibrium flow dynamics in the long wavelength regime (typically for
edge equilibrium flows the local $\rho_*$ is between $1/100$ and $1/30$)
and also expecting $\ddpp\phi > (1/n_e e)\ddpp p_i$ unless the flows are
weak \cite{pet09}.  In conventional tokamaks the low-frequency and
low-beta assumptions referred to above (Eq.\ \ref{eqbpp}) are well
satisfied and the relevant form of MHD is reduced MHD 
\cite{Strauss76,Strauss77}.  In general
(cf.\ Refs.\ \cite{Hahm88a,Brizard07}) 
there are contributions to
polarisation by $\Apl$ but in this limit the terms due to $\Apl$ in $H_2
f$ are small compared to $\Bpp^2/8\pi$, so for these purposes it is
sufficient to keep $\Apl$ only in the parallel kinetic energy term
$mU^2/2$ and in $\Bpp^2/8\pi$.  These considerations are interesting in
their own right and will be treated in a different work.  It is
presently more important, however, to concentrate on the structure of
the theory rather than the details of any particular version, so here
we use the one necessary to obtain familiar reduced MHD forms.  One
further consideration deserves emphasis: no equation beyond the original
statement of $L$ and $H$ requires justification; only $L$ and $H$
themselves.  Once $L$ and $H$ are chosen, all assumption stops and
further results are a matter of derivation.

The Hamiltonian used in Ref.\ \cite{pet09}\ is
\begin{equation}
H = m{U^2\over 2} + \mu B + e\phi - \half m v_E^2
\label{eqmhdh}
\end{equation}
where the square of the ExB velocity and the parallel velocity are
\begin{equation}
v_E^2 = {c^2\over B^2}\abs{\dpp\phi}^2 \qqquad
mU = p_z - {e\over c}\Apl \qqquad \ppzpz{H} = U
\end{equation}
For the derivatives of $H$ we have
\begin{equation}
{\pt H\over\pt\phi} = e
\qquad
{\pt H\over\pt\grad\phi} = - {mc^2\over B^2}\dpp\phi
\qquad
{\pt H\over\pt\Apl} = - {e\over c}U
\qquad
{\pt H\over\pt\grad\Apl} = 0
\end{equation}
and there is no dependence on the Laplacian of either of the field
potentials. 
The functional derivatives are
\begin{equation}
{\delta fH\over\delta\phi} = \int\dW\,\LB fe 
+ {1\over\Bpl}\div\Bpl {fmc^2\over B^2}\dpp\phi\RB
\qqquad
{\delta fH\over\delta\Apl} = - \int\dW\, {e \over c}fU
\end{equation}
The polarisation vector is
\begin{equation}
\vec P = - \rho_M{c^2\over B^2}\dpp\phi
\end{equation}
Inserting these forms into Eq.\ (\ref{eqmomentumb}) using
Eq.\ (\ref{eqfgradh}), we find upon collecting the $\grad V$ terms
\begin{eqnarray}
& & \ptt{}\avg{\rho_M {c\over B^2}\grad\phi\cdot\grad A_\varphi + fp_zb_\varphi}
\nonumber\\ & & \qquad {}
+\pVV{}\avg{\grad V\cdot \LP fp_zb_\varphi\Rdot
- \rho_M{c^2\over B^2}\pvpvp{\phi}\grad\phi
+ {1\over 4\pi}\pvpvp{\Apl}\grad\Apl\RP} = 0
\label{eqmhdtransport}
\end{eqnarray}
which is the desired result.  We have the conserved local toroidal
momentum density, the quantity under the time derivative.  The pieces
are the covariant toroidal angle components of the ExB and parallel
momenta, respectively.  The quantity under the divergence is the radial
flux of the toroidal momentum.  The pieces are the magnetic flutter and
ExB/parallel Reynolds stress, the pure ExB Reynolds stress, and the
Maxwell stress.
This is the result as obtained without the use of any ordering except
that involved in the prescription of $L$.

We also consider this as a mean field theory, in which the flux surface
average is also understood to contain a time average over a mesoscale
range longer than eddy correlation times but shorter than transport
diffusion times.
The fluctuations are assumed to be small amplitude (order $\rho_*$ in
relative amplitude, with velocities normalised to the sound speed).
Following the considerations in Sec.\ \ref{sec:fluxtube}, we apply flute
mode ordering to the fluctuations, with the observed relationship
between $\varphi$ and $\xi$ in covariant components.  Turbulent fluxes
appear in the flux surface averages where the $V$-component of equilibrium
flows is negligible (the magnetic drifts constitute neoclassical
transport, which we neglect here).  In the transported quantity the
contribution due to the fluctuations is neglected.
See Ref.\ \cite{transport} for application of this in a drift-fluid model.
We will also take the single-fluid MHD approximations, where 
each species is assumed to have the same parallel velocity and in the
end there is a single pressure.  In this section, the tilde symbol
denotes fluctuations in the indicated quantities.

To see the ExB toroidal momentum we note that
\begin{equation}
{c\over B^2}\grad\phi\cdot\grad A_\varphi = 
{c\over B^2}\grad\phi\cdot[\grad\varphi\cross(\grad A_\varphi\cross\grad\varphi)]
= R^2{c\over B^2}\vec B\cross\grad\phi\cdot\grad\varphi = \velp
\end{equation}
using Eq.\ (\ref{eqdefb}) and $R^2\abs{\grad\varphi}^2=1$ with $R$ the
toroidal major radius, and noting that the toroidal magnetic field
does not contribute.  
Thus we may combine
\begin{equation}
\avg{\rho_M {c\over B^2}\grad\phi\cdot\grad A_\varphi + fp_zb_\varphi}
= \avg{\rho_M u_\varphi}
\end{equation}
into a toroidal momentum density given by the mass density times
the toroidal flow $u_\varphi$ under the mean field and single-fluid MHD
approximations.  

For the Reynolds and Maxwell stresses, we observe that
\begin{equation}
{c\over B^2}\grad V\cdot\grad\phifl = 
{c\over \chi'{}^2}\LP
{g^{VV}\over g^{VV}g^{\xi\xi}-g^{V\xi}g^{\xi V}}\pVV{\phifl}
+ {g^{V\xi}\over g^{VV}g^{\xi\xi}-g^{V\xi}g^{\xi V}}\pxixi{\phifl}\RP
\end{equation}
Since $F^{\xi V}/B^2=1/\chiv$ the ExB velocity components are
\begin{equation}
\vexfl = {c\over\chi'}\pxixi{\phifl}
\qqquad
\veyfl = - {c\over\chi'}\pVV{\phifl}
\label{eqvfldef}
\end{equation}
The covariant metric coefficients are
\begin{equation}
g_{\xi\xi} = {g^{VV}\over g^{VV}g^{\xi\xi}-g^{V\xi}g^{\xi V}}
\qqquad
g_{\xi V} = { - g^{V\xi}\over g^{VV}g^{\xi\xi}-g^{V\xi}g^{\xi V}}
\end{equation}
Using the index lowering operation
\begin{equation}
\velyfl = g_{\xi\xi}\veyfl  + g_{\xi V} \vexfl 
\end{equation}
we have
\begin{equation}
- {c^2\over B^2}\pxixi{\phifl}\grad V\cdot\grad\phifl = \vexfl\velyfl
\end{equation}
and putting the $2\pi$ normalisation back in,
we have
\begin{equation}
- {c^2\over B^2}\pvpvp{\phifl}\grad V\cdot\grad\phifl = \vexfl\velpfl
\end{equation}
which is the radial/toroidal Reynolds stress.
With magnetic fluctuations given by
\begin{equation}
\bxfl = - {1\over\chi'}\pxixi{\Afl}
\qqquad
\byfl = {1\over\chi'}\pVV{\Afl}
\label{eqbfldef}
\end{equation}
we similarly have
\begin{equation}
{1\over 4\pi}\pvpvp{\Afl}\grad V\cdot\grad\Afl 
= -{B^2\over 4\pi}\bxfl\blpfl
= - \Bfl^V \Bfl_\varphi
\end{equation}
which is the radial/toroidal Maxwell stress.

For the magnetic flutter nonlinearity and the ExB/parallel Reynolds
stress we note that the part of the drifts due to fluctuations is given
by
\begin{equation}
\Rdot = \grad\ptb H\cdot{c\vec F\over eB^2}
= \grad\phifl\cdot{c\vec F\over B^2} - U\grad\Afl\cdot{\vec F\over B^2}
\end{equation}
where to lowest order in $\rho_*$ we approximate $\Bpl\to B$ and neglect
the second order drift term due to $-mv_E^2/2$ in $H$.  The
$V$-component of this is
\begin{equation}
\dot V = {c\over \chi'}\LP\pxixi{\phifl} - U\pxixi{\Afl}\RP
\end{equation}
The flux surface average of the
fluctuation drifts term is then
\begin{equation}
\avg{\ptb fp_zb_\varphi \dot V} \to
\avg{\ptb fp_zb_\varphi{c\over\chi'}\pxixi{\phifl} 
- fUp_zb_\varphi{c\over\chi'}\pxixi{\Afl}}
\end{equation}
These are two different effects in a gyrofluid sense because they
involve different moments of $f$.  In both terms $f\to\ptb f$ 
represents fluctuations because a first-order fluctuation term vanishes
under the time average contained in $\avg{\,}$.  Considering this,
we have the moments
\begin{equation}
\avg{nm\ufl \vexfl b_\varphi + \ptb p_\parallel \bxfl b_\varphi}
\end{equation}
where $\vexfl$ and $\bxfl$ are as given in
Eqs.\ (\ref{eqvfldef},\ref{eqbfldef}).  
These two terms
give the ExB/parallel Reynolds stress and the magnetic flutter
transport pieces.
The distinctions between $U$ and $p_z/m$ and between $p_\parallel$ and
$P_\parallel = p_\parallel + nm\upl^2$ are neglected for small amplitude
fluctuations.  

At the MHD level the sum over species leads back to the total mass
density $\rho_M$ and the total pressure $p$ which replace $nm$ and
$p_\parallel$ (\ie, neglecting anisotropy), respectively.  This leads to
\begin{equation}
\ptt{}\avg{\rho_M u_\varphi} + \pVV{}\avg{
\rho_M \vexfl\velpfl
+ \rho_M\ufl \vexfl b_\varphi 
- {1\over 4\pi} \ptb B^V \ptb B_\varphi
+ \ptb p \bxfl b_\varphi} = 0
\label{eqmeanfield}
\end{equation}
under MHD and mean field approximations having used the ``MHD
Hamiltonian'' given in Eq.\ (\ref{eqmhdh}).  We have placed the two
Reynolds stress terms next to each other and left them separate, for
clarity. 
All of these terms are
easily identifiable with well known processes within the MHD fluid model.
We note that the mean field ordering has been used for evaluation
purposes only, with the actual transport equation for this model given
by Eq.\ (\ref{eqmhdtransport}).

This exercise has served to
prove that the general gyrokinetic toroidal momentum conservation laws
can be brought back to MHD via straightforward application of the MHD
approximations.

\section{The Energy Transport Equation
\label{sec:energy}}

The energy transport equation is much better known, even for gyrokinetic
theory.  However there is an important manipulation concerning the part
of $H$ due to $\phi$.  Decomposing $H=H_0+e\phi+H_P$ as in
Sec.\ \ref{sec:vorticity}, we start with
\begin{equation}
fH = fH_0 + fe\phi + f H_P
\end{equation}
Summing over species and inserting
Eqs.\ (\ref{eqvorticity},\ref{eqpolvector}) we find
\begin{equation}
\sumsp\int\dW\,fH = \sumsp\int\dW\,f(H_0 + H_P) 
+ \div\phi\vec P - \vec P\cdot\grad\phi
\end{equation}
and under the flux surface average
\begin{equation}
\avg{fH} = \avg{f(H_0 + H_P)} - \avg{\vec P\cdot\grad\phi} 
+ \pVV{}\avg{\phi P^V}
\end{equation}
We also have
\begin{equation}
\avg{fH\dot V} = \avg{\LB f(H_0 + H_P)-\vec P\cdot\grad\phi\RB\dot V}
+ \avg{\LP\div\phi\vec P\RP\dot V}
\end{equation}
where as before $\dot V=\Rdot\cdot\grad V$.
Reduction of the time derivative term in Eq.\ (\ref{eqenergycontinuity})
is done the same way as in Sec.\ \ref{sec:momentum}\ for the toroidal
angle derivative term in Eq.\ (\ref{eqmomentumcontinuity}), noting that
the time derivative of $\Bpp^2$ survives, as in the analysis leading to 
Eq.\ (\ref{eqecons}).  We find
\begin{eqnarray}
& & \ptt{}\avg{fH_0 + fH_P - \avg{\vec P\cdot\grad\phi} + {\Bpp^2\over 8 \pi}}
+ \pVV{}\avg{fH_0 \dot V}
\nonumber\\ & & \qquad {}
+ \pVV{}\avg{\LP fH_P - \avg{\vec P\cdot\grad\phi} 
+ \div\phi\vec P\RP\dot V + \ptt{}\phi P^V} 
\nonumber\\ & & \qquad {}
- \pVV{}\avg{ \grad V\cdot\LB
\ptt{\phi}\,\LP f{\pt H\over\pt\grad\phi} 
	- 2\dpp f{\pt H\over\pt\ddpp\phi}\RP
+ \dpp\LP \ptt{\phi}\,f{\pt H\over\pt\ddpp\phi}\RP\RB}
\nonumber\\ & & \qquad {}
- \pVV{}\left\langle \grad V\cdot\LB
\ptt{\Apl}\,\LP {1\over 4\pi}\dpp\Apl + f{\pt H\over\pt\grad\Apl}
	- 2\dpp f{\pt H\over\pt\ddpp\Apl}\RP
\right.
\right.
\nonumber\\ & & \qqquad {} + \left.\left.
\dpp\LP \ptt{\Apl}\,f{\pt H\over\pt\ddpp\Apl}\RP\RB\right\rangle
\nonumber\\ & & \qqquad {}
= 0
\label{eqenergy}
\end{eqnarray}
In practical cases only the terms on the top line are significant.  Under
the time derivative we have the thermal and kinetic energy, ExB energy,
and magnetic energy.  The transport term gives the ExB advection (the
$\grad\phi$ term in $H$ in the drifts) and magnetic flutter (the
$U\grad\Apl$ term in $H$ in the drifts).  All the others are
polarisation and induction corrections.  Inserting the MHD Lagrangian
from Sec.\ \ref{sec:mhd}, and taking the single-fluid 
MHD approximations, we find
\begin{equation}
\avg{fH_0} = \threehalves \avg{p} + \half\avg{\rho_M\upl^2}
\qqquad
\avg{fH_P - \vec P\cdot\grad\phi} = \half\avg{\rho_M v_E^2}
\end{equation}
together with $\Bpp^2/8\pi$ for the energy pieces, and
\begin{equation}
\avg{fH_0\dot V} = \threehalves\avg{\ptb p\vexfl} + \avg{\qepl\bxfl}
\end{equation}
where $\qepl$ is the conductive electron parallel heat flux,
for the dominant transport pieces.  The correspondence with MHD forms is
evident. 
This exercise has served to
prove that the general gyrokinetic energy conservation laws
can be brought back to MHD via straightforward application of the MHD
approximations.

\section{Summary and Discussion}

The main results of this work are the antisymmetric 4-bracket form of
the gyrokinetic equation, Eq.\ (\ref{eqgk4}),
the global conservation laws 
for energy and toroidal momentum,
Eqs.\ (\ref{eqecons},\ref{eqmcons}), 
the local phase space advection equations
for energy and toroidal momentum,
Eqs.\ (\ref{eqlocalenergy},\ref{eqlocalmomentum}), 
the phase space continuity equations
for energy and toroidal momentum,
Eqs.\ (\ref{eqmomentumcontinuity},\ref{eqenergycontinuity}),
and the transport equations for vorticity, toroidal momentum, and
energy,
Eqs.\ (\ref{eqvortrans},\ref{eqmomentumb},\ref{eqenergy})
with Eqs.\ (\ref{eqfgradh},\ref{eqfgradha}) as auxiliaries.  It is
important to note that no ordering assumptions were required to obtain
these.  The only required conditions are the form of the Lagrangian
and Hamiltonian implied by
Eqs.\ (\ref{eqlagrangian},\ref{eqhamiltonian},\ref{eqaction}), the
ability to write the vorticity in the form of a polarisation vector
divergence as in Eqs.\ (\ref{eqvorticity},\ref{eqpolvector}), which
enable the cancellation given in Eq.\ (\ref{eqaplcancel}).  This in turn
leads to the single condition on the form of $H$, that the dependence
upon $\phi$ must be as in Eq.\ (\ref{eqformh}), which leads to
Eqs.\ (\ref{eqvorticity},\ref{eqpolvector},\ref{eqpolform}).  
These results confirm what
Refs.\ \cite{Sugama00,Brizard00}\ already implied and 
Ref.\ \cite{Brizard07}\ already reviewed.  The part of the results that
are novel comprises the 4-bracket form of the gyrokinetic equation, its
use in proving the conservation laws, and that the local forms of the
conservation laws, \ie, the transport equations, have a solid
fundamental basis for any reasonable choice of Lagrangian.  The
correspondence to nonlinear reduced MHD was shown, also without
ordering, in Eq.\ (\ref{eqmhdtransport}).

The usefulness and significance of the ability to arrange the Lie
transforms with which $L_p$ and $H$ are built such that all time and
toroidal angle dependence is kept in the dynamical field variables and
the latter are strictly contained in $H$ (\ie, to obtain the form of
Eq.\ \ref{eqlagrangian}).  Previous work has already shown how to
arrange this, for both the conventional small scale and also for the
newer large scale orderings \cite{Hahm88,Miyato09}.

The role of the cancellation in Eq.\ (\ref{eqaplcancel}) highlights the
role of the vorticity equation in the overall consideration of
momentum.  In a low-frequency fluid drift model, the natural
decomposition of rotation is not poloidal/toroidal but
perp/parallel, with the ExB energy equation controlling the evolution of
the perpendicular flow.  In a gyrokinetic or gyrofluid model, this role
is taken over by the ion gyrocenter density variable, the energy content
is controlled by $fH_E$ (where $H_E$ is the part of $H$ which depends on
$\phi$), and the relevant conserved quantity is, one and the same,
gyrocenter (apparent) charge density and generalised vorticity.  The
appearance of the polarisation density as the divergence of a
polarisation vector is a fundamental property underlying the
conservation laws.  Hence, the local conservation of toroidal momentum
depends on the 
simultaneous conservation of vorticity.  At the level of the equations,
obtaining the one is dependent on the use of the other to provide a
cancellation by which the amplitude of the poloidal magnetic flux (as
opposed to its gradient) is removed from influence on the transport of
toroidal momentum.

The mathematical
functioning of the global conservation equations for energy and
toroidal momentum depends critically on the ability to employ time and
toroidal angle translation symmetry, respectively, in the use of
functional derivatives to evaluate an apparent residual which in fact
vanishes under the integral (the right hand sides of
Eqs.\ \ref{eqlocalenergy},\ref{eqlocalmomentum}).  This in turn depends on
the exact working of the Euler-Lagrange equations for the field
variables, and this in turn is defined in terms of functional derivatives.

At the level of the local equations, these apparent residuals are shown
to be recastable in terms of divergences using techniques similar to the
use of functional derivatives.  Hence, for any $H$, they are recast as
transport fluxes.  All of the transport fluxes which remain significant
under mean field and small fluctuation ordering are found to have clear
MHD analogues, another step in confirming the general solidity of the
theory.  Using a simplified Hamiltonian defined in Eq.\ (\ref{eqmhdh})
which keeps finite gyroradius
residual corrections only in the retention of the ExB energy as a second
order drift, it was shown to be straightforward to recover the MHD limit
of the transport equations (Eq.\ \ref{eqmhdtransport}), 
with the familiar Reynolds and Maxwell
stresses and the magnetic flutter nonlinearity elucidated by a mean
field analysis (Eq.\ \ref{eqmeanfield}).  
This ``MHD Hamiltonian''
provides the link between gyrokinetics and nonlinear reduced MHD at the
total-f level.  The correspondence at the delta-f level was recently
proved elsewhere for gyrofluid theory \cite{braggem}.

A subsidiary analysis was used to show that for small scale fluctuation
ordering, the magnetic flux term $(A_\varphi/c)$, which is formally of
order $\epsilon^{-1}$ as discussed in Ref.\ \cite{Brizard95}\, does not
introduce terms in $H$ at higher order than already are necessary to
evaluate the momentum transport equation.  This result, in
Eq.\ (\ref{eqordering}), is sufficient to allay recent concerns
about the integrity of the treatment of momentum conservation and
transport by gyrokinetics which have been voiced by others
\cite{Catto08}.

Second order drifts are necessary for energetic consistency if
polarisation is nonlinear, but it is possible to have a model with
linearised polarisation and only first order drifts in the gyrokinetic
equation itself and still satisfy energetic consistency.  Such models
also have well behaved local transport equations for toroidal momentum,
as shown in Appendix \ref{app:codes}.  The second order Hamiltonian
terms --- quadratic in the field dependent variables --- must be present
in some form, either as second order drifts in the gyrokinetic equation
or as background field terms, since they control the dynamical energy
accounted for by those variables.

Energetic consistency 
underlies all the derivations, which do not work otherwise.
Any model which does not have this at its heart and
which attempts to go beyond first order field dependence in drift terms
is prescribed to fail on consistency grounds.  It is important to note
that the abovementioned concerns were not done with an energetically
consistent analysis and did not address any of the previous results on
energetic consistency nor even the existence of gyrokinetic field
theory.  However, the seminal references on gyrokinetic field theory
already clearly demonstrated the need to keep fully nonlinear
polarisation at any level of ordering past first order dependence of $H$
upon the field variables \cite{Sugama00,Brizard00}, and this was and is
followed rigorously in field theory treatments of reduced (drift) fluid
equations \cite{Pfirsch96,Pfirsch97}\ and total-f gyrofluid equations
\cite{Strintzi04,Strintzi05}.

Gyrokinetic field theory was not necessary to build the
original gyrokinetic computational models \cite{Lee83,Lee87,Parker93}.  
However, an equivalent field theory can
be built from the given gyrokinetic Poisson equation (multiply by
the time derivative of $\phi$ and find the resulting form with $f$ times
the time derivative of $H$, then use that $H$ to re-derive the equations). 
The above-cited computations and more recent ones derived from them
\cite{ORB,Virginie07,Bottino07,WWang07,McMillan08}\
are consistent with this.
This is why conventional delta-f global computation is on solid
fundamentals.

Gyrokinetic field theory is however the only form of modern gyrokinetic
theory as it is required to provide well founded generalisations in
practically any context.  It is also necessary to any effort which aims
to generalise gyrokinetic theory in practically any way.

{
\bibliography{../paper}
\bibliographystyle{aip}
}

\appendix

\section{Use of Functional Derivatives}

The Euler-Lagrange equations for the fields are found by identifying the
functional derivatives of $L$ with respect to each field potential and
setting them to zero.  For an introduction to the use of functional
derivatives in this type of field theory see Section II B
of the review by Morrison \cite{Morrison05}.
For background on functional derivatives within the general
context of the calculus of variations see the text by
Gelfand and Fomin \cite{GelfandFomin}.
For the more familiar version by which stress tensor theorems are proved
within classical field theory
see the text by Landau and Lifshitz \cite{LandauLifshitz}.

The functional derivatives are defined with
respect to the space covered by $\dV$, so that, e.g., variation of
$\phi$ results in
\begin{equation}
\delta L(\delta\phi) \equiv \int\dV\,\delta\phi\,{\delta L\over\delta\phi}
\end{equation}
in which it is understood that the definition of the functional
derivative includes velocity space integration.
The arbitrariness of $\delta\phi$ and the extremal requirement for $L$
give the field equation as 
\begin{equation}
{\delta L\over\delta\phi} = 0
\end{equation}
Since the dependence of $L$ on $\phi$ is through $H$ only then this is
equivalent to 
\begin{equation}
\sumsp{\delta fH\over\delta\phi} = 0
\end{equation}
Variation of $\Apl$ results in
\begin{equation}
\delta L(\delta\Apl) \equiv \int\dV\,\delta\Apl\,{\delta L\over\delta\Apl}
\end{equation}
and the arbitrariness of $\delta\Apl$ and the extremal requirement for $L$
give the field equation as 
\begin{equation}
{\delta L\over\delta\Apl} = 0
\end{equation}
Since the dependence of $L$ on $\Apl$ includes the field term 
$-\Bpp^2/8\pi$ this is equivalent to 
\begin{equation}
\sumsp{\delta fH\over\delta\Apl} = {1\over 4\pi}\ddpp\Apl
\end{equation}
These are the field equations for arbitrary $H$ given the field term
$\Bpp^2/8\pi$.  The induction equation depends on the form of this term;
for example the version of $\Bpp^2$ 
which gives a result closer to the form of
Amp\`ere's law used in the Grad-Shafranov equation for MHD equilibrium is
\begin{equation}
\Bpp^2 = {1\over R^2}\abs{\dpp(\Apl R)}^2
\end{equation}
where $R$ is the toroidal major radius.  In this case the field equation
is given by
\begin{equation}
\sumsp{\delta fH\over\delta\Apl} 
= {R\over 4\pi}\div{1\over R^2}\dpp(\Apl R)
\end{equation}
with the new form on the right side.
Due to the axisymmetry of $R$, either of these
forms may be used in the above derivations.

\subsection{Integrals over $f$ and $H$}

The conservation laws depend on phase space integrals over expressions
of $f$ and $H$ in the form
\begin{equation}
\sumsp\int\dL\,f\,d_a\,H
\end{equation}
where $d_a$ is a differential operator with the linearity property, 
(e.g., a time derivative).  In turn, the dependence of $H$ on the fields
involves the field amplitudes and their derivatives which we can
schematically express as
\begin{equation}
H=H(\phi,d_b\phi)
\end{equation}
where $d_b$ is another differential operator with either the linearity
or the the Hermitian property (e.g., $\grad$ or $\ddpp$ or $J_0$).  
The relevant differential of $H$ is
\begin{equation}
d_a H = {\pt H\over\pt\phi} d_a\phi 
+ {\pt H\over\pt d_b \phi} d_a d_b\phi 
\end{equation}
We can turn the integral into a functional derivative form if and only
if the two operators commute:
\begin{equation}
\int\dL\,f\,d_a\,H = \int\dL\,{\delta fH\over\delta\phi}d_a\phi
\qquad\iff\quad (d_a d_b - d_b d_a)\phi = 0
\end{equation}
If all the $d_b$ in $H$ commute with $\ppt{}$ then we have time symmetry
which is a prerequisite for energy conservation.
If all the $d_b$ in $H$ commute with $\pt/\pt\varphi$ then we have
toroidal angle (axi-)symmetry
which is a prerequisite for toroidal momentum conservation.  In general
the commutator does not vanish if $d_a$ is $\grad$, since $d_b$ involves
$B$ which is spatially dependent.  In stellarator geometry 
$\pt B/\pt\varphi\ne 0$ and therefore toroidal momentum conservation is
not conserved (it is exchanged with $\vec B$).  In tokamak geometry
poloidal momentum is exchanged with $\vec B$ but toroidal momentum is
conserved in the ideal case (no destruction of axisymmetry).

\section{Equivalent Hamiltonians to Conventional Models}

As an example of how one can back-construct a gyrokinetic field theory
from a conventional model, we consider the gyrokinetic Poisson equation
as currently used in most numerical simulations 
\cite{ORB,Virginie07,Bottino07,McMillan08}.  The gyrocenter charge
density piece is kept with gyroaveraging and the polarisation is given
by a background ion density,
\begin{equation}
\sumsp\int\dW\LB eJ_0 f \RB + \div{n_0 M_i c^2\over B^2}\dpp\phi = 0
\label{eqorbpol}
\end{equation}
For this to come from a Lagrangian it has to be the functional
derivative with respect to $\phi$.  We may rebuild this by multiplying
by $\delta\phi$ and integrating over the volume, so that
\begin{equation}
\sumsp\int\dL\,\delta\phi\LB eJ_0 f \RB + 
\int\dV\,\delta\phi\, \div{n_0 M_i c^2\over B^2}\dpp\phi = 0
\end{equation}
Here, all we need know about $J_0$ is that it is Hermitian.
Doing the relevant integrations by parts we have
\begin{equation}
\sumsp\int\dL\LB f eJ_0 (\delta\phi) \RB - 
\int\dV\, {n_0 M_i c^2\over B^2}\dpp\phi\cdot\dpp(\delta\phi) = 0
\end{equation}
We identify the relevant parts of the Lagrangian as
\begin{equation}
L = \cdots - \sumsp\int\dL\,f(eJ_0\phi) 
+ \int\dV\, {n_0 M_ic^2\over 2B^2}\abs{\dpp\phi}^2
\end{equation}
Variation of this with respect to $\phi$ indeed recovers
Eq.\ (\ref{eqorbpol}).  Inspection of the version of the gyrokinetic
equation used in these references shows that indeed $H_E=eJ_0\phi$ is
used as the perturbed Hamiltonian in the drifts.  Therefore, energetic
consistency is assured.  

Most versions of this model are
considered with electrons in adiabatic force balance parallel to the
magnetic field,
\begin{equation}
\int\dW f_e = n_0\LB 1 + {e_0\over T_e}\LP \phi-\avg{\phi}\RP\RB
\label{eqadiabatice}
\end{equation}
with gyroaveraging neglected and with the same $n_0$ as in the
background ExB energy term.  
The factor $e_0$ is the fundamental
unit of charge, so written to distinguish from use of $e$ as a species
charge.
The temperature $T_e$ is a flux function and
the subtraction of the flux surface average reflects the vanishing of
the parallel gradient for this component.  This state of force balance
is expected to evolve adiabatically in local thermodynamic equilibrium,
so the contribution to $L$ is a profile anchor piece plus a fluctuation
energy piece,
\begin{equation}
L_{{\rm electrons}} = n_0e_0\phi + n_0{e_0^2\over 2T_e}\LP \phi-\avg{\phi}\RP^2
\end{equation}
which becomes a field term.
With this substitution made in $L$, the polarisation equation is
\begin{equation}
n_0{e_0^2\over T_e}\LP\phi-\avg{\phi}\RP 
- \div{n_0 M_i c^2\over B^2}\dpp\phi
= - n_0 e_0 + \sumions\int\dW\, eJ_0 f 
\end{equation}
where the sum is over the ions only (for singly charged ions the same
$n_0$ is used everywhere; for more species the appropriate adjustments
are made to keep the profiles charge neutral).   The term $n_0 e_0$
subtracts the profile piece from the ions.
Energy conservation for this model was proved for the ORB code
in Ref.\ \cite{ORB}.
The same model is also used by the GYSELA code \cite{Virginie07}.

The form given by Lee in the original gyrokinetic/Poisson system 
\cite{Lee83,Lee87,Federici87} is 
\begin{equation}
\Psi = J_0\phi - {q\over 2T}{v_t^2\over\Omega_i^2}
\abs{{\pt\phi(\vec R)\over\pt\vec R}}^2
\label{eqleepsi}
\end{equation}
where $\Psi$ is the potential with which particles are pushed in a
gyrokinetic particle in cell model,
$\phi$ is given strictly as a function of the gyrocenter position
$\vec R$, 
the symbol used for gyroaveraging is replaced by our use of $J_0$,
and $q$, $T$, and $(v_t/\Omega_i)^2=mTc^2/q^2B^2$ are constant
parameters, with $q$ the same as our $e$.  This is Eq.\ (2) of
Ref.\ \cite{Lee87}.  It is also an electrostatic model with $\Apl=0$ and
$p_z/m$ interchangeable with a parallel velocity.

In our notation we write this as
$H_E=e\Psi$ with
\begin{equation}
H_E = eJ_0\phi - {mc^2\over 2B^2}\abs{\grad\phi}^2
\end{equation}
In other words, this is the same choice as
our MHD Hamiltonian (Eq.\ \ref{eqmhdh}), 
except for the use of the gyroaveraged $\phi$ as the first term and the
terms dependent upon $\Apl$ in the MHD case.
Lee's gyrokinetic Poisson equation is written in Eq.\ (3) of
Ref.\ \cite{Lee87} as
\begin{equation}
\delsq\phi - {\tau\over\lambda_D^2}(1-\Gamma_0)\phi
+{\rho_s^2\over\lambda_D^2}\div{n_i-n_0\over n_0}\dpp\phi 
= -4\pi e_0(\overline{n}_i - n_e)
\label{eqleepol}
\end{equation}
The operator $\Gamma_0$ results from $J_0^2$ integrated against a
Maxwellian and is given formally by 
multiplication in wavenumber space of Fourier coefficients by
the $\Gamma_0(b)=e^{-b}I_0(b)$, where $I_0$ is the zeroth modified
Bessel function, and the argument is 
$b=\kkpp\rho^2$ with $\rho$ the
species thermal gyroradius given by $\rho^2=mTc^2/e^2B^2$.
The factor $\overline{n}_i$ with the
overbar denotes velocity space integration of $J_0 f_i$, the $n_i$ without
the overbar is velocity space integration of $f_i$, 
and $\tau/\lambda_D^2=4\pi n_0 e_0^2/T_i$
and $\rho_s^2/\lambda_D^2=4\pi n_0M_ic^2/B_0^2$ 
and $n_0$ are constant parameters.  
Using the definitions in his Eqs.\ (4-7), this is found to be equivalent
to 
\begin{equation}
{1\over 4\pi}\delsq\phi + \sumsp\int\dW\LB
F^M{e^2\over T}(J_0^2-1+\rho^2\kkpp)\phi
+{1\over\Bpl}\div\Bpl{fmc^2\over B^2}\dpp\phi + eJ_0 f\RB = 0
\end{equation}
where $F^M$ is a Maxwellian with species parameters $n$ and $T$, 
the correction factor including $J_0^2$ and $\kkpp\rho^2$
restores full finite gyroradius (FLR) effects to the field term,
the sum 
over species using $m_e\to 0$ includes the electrons only in the last
term $eJ_0 f$, and we have restored a toroidal model with the factors
$B$ and $\Bpl$ arranged to preserve Hermicity of all the operators.
The first term is true charge separation, the second
with $F^M$ is a field term, and the rest is the part due to the
dependent variable.  If we prescribe an electrostatic Lagrangian,
\begin{equation}
L = \sumsp\int\dL\,f\,L_p
+ \sumsp\int\dL\,F^M\LBR {e^2\over 2T}\LB\phi^2-(J_0\phi)^2\RB
- {mc^2\over 2B^2}\abs{\dpp\phi}^2\RBR
+ \int\dV\,{1\over 8\pi}\abs{\grad\phi}^2
\end{equation}
with the particle Lagrangian given by
\begin{equation}
L_p = (\vec A+p_z\bunit)\cdot\Rdot
+ {mc\over e}\mu\thetadot - \LP{p_z^2\over 2m}+\mu B+H_E\RP
\label{eqlp}
\end{equation}
then the Euler-Lagrange equations recover gyrokinetic equations with
$e\Psi$ as the $\phi$-dependent part of $H$ and the polarisation
equation given by Lee as his gyrokinetic Poisson equation;
in other words, Eqs.\ (1-3) of Ref.\ \cite{Lee87}.  
These were given for slab geometry but
with the dependences of $B$ placed as shown here together with the use
of $B_0$ where indicated, this recovers a toroidal model with some extra
field terms in $L$ compared to ours.
Since the field correction term in the polarisation equation
is the functional derivative of a
positive definite quantity, it still conserves energy.  Since we are
able to recast the field terms as a field Lagrangian, and to show
otherwise that the
particle pushing potential and the gyrocenter charge terms arise from
the same term $H_E$ in the Hamiltonian,
energetic consistency is entirely satisfied.

Various simplified versions are given.  For example, in the GTC code
\cite{ZLin95,ZLin98}\ the particle pushing potential is consistent with
$H_E=eJ_0\phi$ and the gyrokinetic Poisson equation is given as
\begin{equation}
{\tau\over\lambda_D^2}(1-\Gamma_0)\phi = 4\pi e_0(\overline{n}_i - n_e)
\end{equation}
then we have a linearised polarisation model consistent with
\begin{equation}
L = \sumsp\int\dL\,f\,L_p
+ \sumsp\int\dL\,F^M{e^2\over 2T}\LB\phi^2-(J_0\phi)^2\RB
\end{equation}
where $L_p$ is as in Ref.\ (\ref{eqlp}) with the simplified $H_E$,
and adiabatic electrons are
prescribed through Eq.\ (\ref{eqadiabatice})
above.
Since using this $L$ the above polarisation equation is recovered along
with use of $\Psi=J_0\phi$ in the gyrocenter drifts, the model remains
energetically consistent.  Approximation of the $F^M(J_0^2-1)$ to long
wavelength then recovers the ORB/GYSELA model mentioned above.

Lee mentions these models also in Ref.\ \cite{Lee83}, giving their
source in another model used to obtain his
Eqs.\ (21,22).  The particle pushing potential is given as
\begin{equation}
\Psi = J_0\phi+{e\over 2T}\LB \LP J_0\phi\RP^2-J_0(\phi^2)\RB
\end{equation}
in our notation.  The factor $1/T$ is obtained by approximating $\pt
f/\pt\mu$ as $(-B/T)f$ since the second order term leads to the
polarisability, which should be dominated by the largest scale part of
$f$ which should not depart significantly from a Maxwellian.  This
$\Psi$ was then used to recover the model given in 
Eqs.\ (\ref{eqleepsi},\ref{eqleepol})
by using the long
wavelength approximation of the second order term.  We write the
interaction part, $fH_E=fe\Psi$, of the Lagrangian as 
\begin{equation}
L = \cdots - \sumsp\int\dL\,f\,H_E
\end{equation}
then undo this effective integration by parts under the phase space
integration to obtain a similar model with
\begin{equation}
H_E = eJ_0\phi - {e^2\over 2B}\pmumu{}\LB J_0(\phi^2)-\LP J_0\phi\RP^2\RB
\end{equation}
With this form the field term FLR correction is unnecessary and with
quasineutrality we recover use of an $L$ in which the only dependence
upon $\phi$ is in $H$.
The corresponding polarisation equation is
\begin{equation}
\sumsp\int\dW\LB eJ_0 f + \LP J_0\magnet J_0 - [J_0\magnet]\RP\phi\RB = 0
\end{equation}
where the polarisability $\magnet$ is given by
\begin{equation}
\magnet = -{e^2 \over B}\pmumu{f}
\end{equation}
It is very difficult to be able to take this derivative in a particle in
cell model, which is why this second order term is usually approximated
in some fashion.

This is very close to the version given by Hahm \cite{Hahm88}, who
obtains an additional term from the second order part of the Lie
transform so that
\begin{equation}
H_E = e\avg{\phi}\subR - {e^2\over 2B}\pmumu{}\avg{\ptb\phi^2}\subR
- {mc^2\over 2B^2}\avg{\grad\ptb\Phi\cdot\bunit\cross\grad\ptb\phi}\subR
\end{equation}
where in this case the subscripted angle brackets denote $J_0$ at the
particle level and
\begin{equation}
\ptb\phi = \phi-\avg{\phi}\subR \qqquad
\ptb\Phi = \int^\vartheta\ptb\phi\,d\vartheta
\end{equation}
The first of the two second order terms is equivalent to Lee's.  The
second involves an indefinite gyrophase integral and is difficult to
compute.  The equivalent polarisation equation is then given
keeping all the
relevant terms through use of the Lie back-transform.  As noted above,
this is equivalent to the use of field theory \cite{Brizard07}.
However, the added term in $H_E$ yields a contribution to 
the polarisation equation which is one order down
in $\rho_*$ from the others, and it is never kept in computations.
Nevertheless, with the approximations always effectively made in $L$ and
nowhere thereafter, exact energetic consistency is preserved in all of
the versions.
This effectively brings us back to Lee's $H_E$.

\subsection{Momentum conservation with linearised polarisation
\label{app:codes}}

Here we demonstrate the existence of a familiar toroidal momentum
conservation equation within the simplest possible form of a global
gyrokinetic model.  The dynamics is electrostatic, quasineutral,
gyroaveraging is neglected, and polarisation is provided by a background
field term.  The Lagrangian is
\begin{equation}
L=\sumsp\int\dL\, f\,L_p + \int\dV\,{\rho_Mc^2\over 2B^2}\abs{\dpp\phi}^2
\end{equation}
where $\rho_M$ is given by a static profile (or constant), 
and $L_p$ is of the form given in Eq.\ (\ref{eqlagrangian}), with
\begin{equation}
H = {p_z^2\over 2m} + \mu B + e\phi
\end{equation}
The corresponding polarisation equation is 
\begin{equation}
\div{\rho_Mc^2\over B^2}\dpp\phi + \sumsp\int\dW\,ef = 0
\label{eqsimplepol}
\end{equation}
This is the same model as in the ORB code mentioned above, except we
simplify by neglecting $(1-J_0)$ hence FLR effects.

The particle drifts are given by Eq.\ (\ref{eqdriftmotion}), with
spatial part
\begin{equation}
\Bpl\Rdot = \grad\phi\cdot{c\vec F\over B} 
+ \mu\grad B\cdot{c\over e}{\vec F\over B}
- {p_z^2\over m}\LP\div{c\over e}{\vec F\over B}\RP
+{p_z\over m}\vec B
\end{equation}
with the pieces
identified as the ExB velocity, the grad-B and curvature drifts, and the
parallel velocity.  The radial component of this is then
\begin{equation}
\dot V = \grad V\cdot\Rdot = (v_E)^V + (v_{\grad B})^V + (v_C)^V
\end{equation}
which we use below (the parallel piece does not contribute).

The steps to the momentum equation are the derivations of
Eqs.\ (\ref{eqlocalmomentum},\ref{eqmomentumcontinuity},\ref{eqvortrans}) 
which remain unchanged, and the polarisation vector 
\begin{equation}
\vec P = - {\rho_Mc^2\over B^2}\dpp\phi
\end{equation}
with the only difference to the MHD model at this point being the 
static $\rho_M$.  Specifically, we still have
\begin{equation}
\ptt{}\avg{fe} = - \avg{\div\ptt{}{\rho_Mc^2\over B^2}\dpp\phi}
= -\pVV{}\avg{\ptt{P^V}}
\end{equation}
with the background $\rho_M$.  The cancellation in
Eq.\ (\ref{eqaplcancel}) remains intact, and we have
Eq.\ (\ref{eqmomentumb}).  We evaluate
\begin{equation}
\avg{fe\pvpvp{\phi}} = - \pVV{}
\avg{\rho_M{c^2\over B^2}\pvpvp{\phi}\grad V\cdot\grad\phi}
\end{equation}
to obtain
\begin{equation}
\ptt{}\avg{\rho_M(v_E)_\varphi+fp_zb_\varphi}
+\pVV{}\avg{\grad V\cdot\LP fp_zb_\varphi\Rdot
-\rho_M{c^2\over B^2}\pvpvp{\phi}\grad\phi\RP} = 0
\end{equation}
as the local toroidal
momentum conservation equation, \ie, the toroidal
momentum transport
equation.  The pieces are easily identified as the ExB and parallel
components of the toroidal momentum, the ExB/parallel Reynolds stress,
and the toroidal component of the pure ExB Reynolds stress, all familiar
effects.  In the drifts, $(v_E)^V$ gives the ExB/parallel Reynolds
stress, while $(v_{\grad B})^V$ and $(v_C)^V$ give the grad-B and
curvature drifts which comprise the neoclassical transport
(to evaluate these, of course, one needs treatment of the collisions;
cf.\ the review by Hinton and Hazeltine \cite{HintonHazeltine76}).
Except for the corrections arising through $J_0$, this is the
toroidal momentum equation satisfied by the model used in the 
conventional numerical models mentioned above.

The recent NEMORB code \cite{Bottino09}\ is the electromagnetic version
of ORB.  It reestablishes $\Apl$ at first order in $H$, placing the
second order term alongside the polarisation field term.  The version 
neglecting FLR effects is given by
\begin{equation}
L=\sumsp\int\dL\, f\,L_p 
+ \int\dV\LP
{\rho_Mc^2\over 2B^2}\abs{\dpp\phi}^2 
- {n_0e^2\over 2Mc^2}\Apl^2 - {1\over 8\pi}\abs{\dpp\Apl}^2
\RP
\end{equation}
with reduced mass $M=m_eM_i/(m_e+M_i)$ and with
$L_p$ as in Eq.\ (\ref{eqlagrangian}), and $H$ given by
\begin{equation}
H = {p_z^2\over 2m} + \mu B + e\LP\phi-{p_z\over mc}\Apl\RP
\end{equation}
The polarisation equation is unchanged, given by
Eq.\ (\ref{eqsimplepol}), and the induction equation is
\begin{equation}
\LP{\omega_p^2\over c^2}-\ddpp\RP\Apl 
= \sumsp\int\dW\,{4\pi e\over mc}\,p_z\,f
\qqquad
\omega_p^2 = {4\pi n_0 e^2\over M}
\end{equation}
These are the simplifications of Eqs.\ (14,25) of Ref.\ \cite{Hahm88a},
respectively, descending accordingly 
from the simplification of $H$ (see the
discussion at the beginning of Sec.\ \ref{sec:mhd}).

The corresponding toroidal momentum transport equation is
\begin{equation}
\ptt{}\avg{\rho_M(v_E)_\varphi+fp_zb_\varphi}
+\pVV{}\avg{\grad V\cdot\LP fp_zb_\varphi\Rdot
-\rho_M{c^2\over B^2}\pvpvp{\phi}\grad\phi
+{1\over 4\pi}\pvpvp{\Apl}\grad\Apl\RP
} = 0
\end{equation}
which merely adds the Maxwell stress as the last term.  Other terms from
the field equations are put into the form $\pt S/\pt\varphi$ with scalar
$S$ and are annihilated by the flux surface average.
The neglect of
FLR effects in this demonstration is merely for clarity; we may restore
$J_0$ and expand it in a series of Laplacians to recover several
correction effects following the procedure for Laplacian field variable 
dependence given in Sec.\ \ref{sec:laplacians}.  Hence, we have
demonstrated that momentum conservation and transport in ORB and NEMORB,
as well as other related codes, is on a solid foundation.

\section{Torque due to a Charge Source}

In exceptional cases the particle sources are not in charge balance.
The exemplary case is ion orbit loss \cite{Shaing89,Heikkinen98}, 
in which the ions on large trapped
(banana) orbits impact material surfaces while the electrons remain
confined.  Strictly speaking this is a transport effect and is accounted
for by the drifts term and a loss flux through the boundary.  In
practice, however, it is modelled by a localised 
loss term in the ion continuity
equation.  The corresponding vorticity transport equation is
\begin{equation}
\ptt{}\avg{\Omega} - \pVV{}\avg{fe\dot V} = \avg{S_\Omega}
\qqquad S_\Omega = -e_i\left.\dtt{f_i}\right\vert_{{\rm loss}}
\end{equation}
Since $S_\Omega$ is a scalar quantity we may specify
\begin{equation}
\ddpp s = S_\Omega
\end{equation}
with $s=\ppV{s}=0$ on the magnetic axis ($V=0$), without loss of generality.
With $\div\vec P= - \Omega$ we have
\begin{equation}
\pVV{}\avg{\ptt{P^V} + fe\dot V - \grad V\cdot\grad s} = 0
\end{equation}
so that the flux surface average quantity vanishes as before.

In the toroidal momentum continuity equation the terms which involve
$A_\varphi$ are
\begin{eqnarray}
& & \ptt{}{A_\varphi\over c}\avg{fe} + \pVV{}{A_\varphi\over c}\avg{fe\dot V}
+ {A_\varphi\over c}\avg{S_\Omega}
= \ptt{}\avg{-{1\over c}\vec P\cdot\grad A_\varphi}
 - \avg{{1\over c}\grad s\cdot\grad A_\varphi}
\nonumber\\ & & \qquad {} 
-\pVV{}{A_\varphi\over c}\avg{\ptt{P^V} + fe\dot V + \grad V\cdot\grad s}
\end{eqnarray}
The terms on the last line vanish.  With the manipulations which find
\begin{equation}
\grad s\cdot\grad A_\varphi 
= R^2\grad\varphi\cross(\grad s\cross\grad\varphi)\cdot\grad A_\varphi
= R^2\vec B\cross\grad s\cdot\grad\varphi
\end{equation}
we find
\begin{equation}
\ptt{}{A_\varphi\over c}\avg{fe} + \pVV{}{A_\varphi\over c}\avg{fe\dot V}
+ {A_\varphi\over c}\avg{S_\Omega}
= \ptt{}\avg{-{1\over c}R^2\vec B\cross\vec P\cdot\grad\varphi}
 - \avg{{1\over c}R^2\vec B\cross\grad s\cdot\grad\varphi}
\end{equation}
In the case of the MHD Hamiltonian $\vec P = -\rho_M(c^2/B^2)\dpp\phi$
and the first term gives the ExB covariant toroidal momentum (equivalent
to toroidal angular momentum).  Hence the second term can be represented
as a toroidal torque
\begin{equation}
\vec T = {1\over c}\vec B\cross\grad s
\qqquad
R^2\vec T\cdot\grad\varphi = T_\varphi
\end{equation}
and the vorticity source is given by
\begin{equation}
S_\Omega = \div{c\over B^2}\vec T\cross\vec B
\end{equation}

\end{document}